\documentclass[aip,reprint]{revtex4-1}

\usepackage{caption}
\usepackage{subcaption}
\usepackage[boxed,commentsnumbered]{algorithm2e} 
\usepackage{comment}

\usepackage{epsfig}

\newcommand{\ldir}{mysections/fig/extra}
\newcommand{\figJac}{mysections/fig}

\newtheorem{theorem}{Theorem}[section]

\newtheorem{definition}[theorem]{Definition}

\newtheorem{observation}{Observation}

\def\IB{{\it bottom}}
\def\IT{{\it top}}

\newcommand{\Cp}{Cayley point}
\newcommand{\Cr}{Cartesian realization}

\newcommand{\comm}[1]{}

\newcommand{\helix}{molecular unit}

\newcommand{\acg}{G}

\newcommand{\aU}{atom marker}
\newcommand{\chartt}{chart}
\newcommand{\atlas}{atlas}

\newcommand{\defref}[1]{Definition~\ref{#1}}

\newcommand{\figref}[1]{Fig.~\ref{#1}}

\draft 

\begin{document}


\title{\Large\bf Best of Both Worlds: Uniform sampling in Cartesian and Cayley Molecular Assembly Configuration Space} 



\author{Aysegul Ozkan}
\email{aozkan@cise.ufl.edu}

\author{Meera Sitharam}
\email{sitharam@cise.ufl.edu}
\affiliation{CISE department, University of Florida, CSE Bldg, Gainesville,
FL 32611-6120 This research was supported in part by
NSF Grants DMS- 0714912,
and  CCF-1117695}


\date{\today}

\begin{abstract}
    EASAL (efficient atlasing and sampling of assembly landscapes)
        is a recently reported geometric method for
            representing, visualizing, sampling and computing integrals over
                the potential energy landscape tailored for
                    small molecular assemblies.
                        EASAL's efficiency arises from the fact that small
                        assembly landscapes
                            permit the use of so-called Cayley parameters
                            (inter-atomic
                                distances) for geometric representation and
                                sampling of the assembly
                                    configuration space regions; this
                                    results in their
                                        isolation, convexification,
                                            customized sampling  and
                                            systematic
                                                traversal using a
                                                comprehensive topological
                                                    roadmap, 
 ensuring  reasonable coverage of crucial but
     narrow regions of low effective dimension.

    However,
    this alone is inadequate for accurate computation of 
    configurational entropy and other integrals, required for estimation of 
    both free energy and kinetics - 
    where it is essential to obtain \emph{uniform} sampling in 
    appropriate cartesian or moduli space parameterization.
    Standard adjustment of Cayley 
    sampling via the Jacobian of the map between the 
    two parameterizations is fraught with challenges stemming from
    an illconditioned Jacobian.

    This paper formalizes and analyzes these challenges to provide
    modifications to EASAL that secure the advantages 
    of Cayley sampling while ensuring 
    certain minimum distance and coverage relationships between sampled
    configurations - in Cartesian space.
    The modified EASAL's performance is compared with the
    basic EASAL and the data are presented for
    Human and Rat Islet Amylin Polypeptide (HiAPP, PDB-2KJ7 and RiAPP
    PDB-2KB8) dimerization 
    (the two differ in only 6 out of 37 residues, but the 
    former aggregates into fibrils, while the latter does not).  

\end{abstract}


\maketitle 


\section{Introduction}
\label{intro}
Understanding and engineering a variety of supramolecular assembly,
packing and docking processes even for small assemblies requires 
a comprehensive atlasing of the topological roadmap of the
constant-potential-energy regions, as well as the ability to isolate and
sample such regions and their boundaries even if they are narrow and
geometrically complex. 
A recently reported geometric  method called 
EASAL (efficient atlasing and sampling of assembly landscapes) 
\cite{Ozkan2014MainEasal} 
provides such comprehensive atlasing as well as customized and efficient sampling of
its regions, crucially employing so-called Cayley or distance parameters.   
However,  for developing hybrids that combine the complementary strengths
of EASAL and prevailing methods  that  predict
noncovalent binding affinities and kinetics, 
accurate computation of 
configurational entropy and other integrals is essential. This in turn
requires uniform distribution over the cartesian or appropriate Cartesian space
parametrization. Standard adjustments using the Jacobian of the
Cartesian-Cayley map poses multiple challenges due 
to illconditioning.  
This paper analyzes these challenges and develops a modification of EASAL
that combines the best of both worlds  of Cayley sampling with uniform
distribution in Cartesian space.

\subsection{Recent Related Work}
\label{related}
A number of very recent results 
are directly related to or build upon the approach presented here. 
First, the basic EASAL approach was first described in \cite{Ozkan2011}. 
The approach is discussed in detail in \cite{Ozkan2014MainEasal}, which gives 
EASAL-based 
computations of entropy integrals for clusters of assembling 
spherical particles 
that  both simplify and extend 
the methodology and computational results of 
\cite{Holmes-Cerfon2013} that were reported after \cite{Ozkan2011} appeared.
A multi-perspective comparison of variants of EASAL including the 
modification described here with traditional Montecarlo sampling of 
the assembly landscape of 2
transmembrane helices  is given in
\cite{Ozkan2014MC}   
with a view towards leveraging complementary strengths for hybrid methods. 
An application of EASAL towards detecting assembly-crucial
inter-atomic interactions for viral capsid self-assembly 
is given in \cite{Wu2012, Wu2014Virus} 
(applied to 3 viral systems - Minute virus of Mice (MVM), Adeno-associated virus (AAV), and Bromo-mosaic
virus (BMV)). 
Finally the architecture and functionalities of an 
opensource software implementation of the basic EASAL is described
in \cite{Ozkan2014TOMS}.

\subsection{Previous Work, Scope and Motivation}
There has been a
long and distinguished history of configurational entropy and free energy
computation methods
\cite{kaku, Andricioaei_Karplus_2001,
    Hnizdo_Darian_Fedorowicz_Demchuk_Li_Singh_2007, 
        Hnizdo_Tan_Killian_Gilson_2008,
            Hensen_Lange_Grubmuller_2010,
                Killian_Yundenfreund_Kravitz_Gilson_2007,
                    Head_Given_Gilson_1997,
                    GregoryS201199, doi:10.1021/jp2068123}, many of which
                    use as input the
                    configuration trajectories of Molecular Dynamics or
                    Monte Carlo
                    sampling which are known to be nonergodic, whereby
                    locating and isolating
                    narrow channels and their boundaries, i.e., regions of
                    low effective
                    dimension separated by high energy barriers might take
                    arbitrarily long, requiring several
                    trajectories starting from different initial
                    configuations.
                    
This also causes problems for many entropy computation methods that rely
 on principal component  analyses of the covariance matrices  from a
     trajectory  of MC samples in internal coordinates,  followed by a
            quasiharmonic
                         \cite{Andricioaei_Karplus_2001}
                                                or nonparametric (such as
                                                nearest-neighbor-based)
                                                                                      \cite{Hnizdo_Darian_Fedorowicz_Demchuk_Li_Singh_2007}
                                                                                      estimates.
 Since MC trajectories are not geometrically optimized, these
         methods are generally known to
                \emph{overestimate} the
                       volumes of
                          configuration space
                                regions with high
                                geometric or topological complexity,
                                    even when hybridized with higher order
                                    mutual information
                                        \cite{Hnizdo_Tan_Killian_Gilson_2008},
                                                and
                                                    nonlinear kernel
                                                        methods,  such as
                                                        the
                                                             Minimally
                                                             Coupled
Subspace approach
        of
           \cite{Hensen_Lange_Grubmuller_2010}.
Most of the above methods do not explicitly restrict the number of atoms 
in each of the assembling rigid molecular components,
and in fact they are used for assembly or folding. 
For cluster assemblies from spheres,  
(with $k\le 12$), there are a number of 
methods \cite{Holmes-Cerfon2013, Hagen1993, Doye1996, Meng2010, Gazzillo2006}
to compute free energy and configurational entropy for 
subregions of the configuration space, and some of these subregions are 
the entire configuration spaces of small molecules such as cyclo-octane
\cite{Martin2010, Jaillet2011, Porta2007}. 
These include robotics and computational geometry based methods 
such as \cite{GregoryS201199} ($n=3$).
These methods are used to give bounds or to approximate
configurational entropy without relying on Monte Carlo or
Molecular
Dynamics sampling. 
Note that there is 
an extensive literature purely on computing minimum potential
energy configurations: these are 
are not relevant to this paper; neither are 
simulation-based methods for
free-energy computation of large assemblies
starting from known free energy values and formation rates 
for assembly intermediates formed from a small number of subassemblies.

Essentially, even  for small assemblies, 
barring a few exceptions such as
 \cite{Holmes-Cerfon2013}, \cite{Porta2007},
\cite{Yao_Sun_Huang_Bowman_Singh_Lesnick_Guibas_Pande_Carlsson_2009}, and
\cite{Gfeller_DeLachapelle_DeLos_Rios_Caldarelli_Rao_2007,Varadhan_Kim_Krishnan_Manocha_2006,
Lai_Su_Chen_Wang_2009, 10.1371/journal.pcbi.1000415}, 
most prevailing methods do not extract a 
high-level, topological roadmap
of the boundary relationships between the constant-potential-energy regions. 
Similarly,  most prevailing methods of sampling and volume computation 
     are not \emph{explicitly} tailored or specialized 
 to leverage this relative geometric simplicity of constant-potential-energy
 regions of assembly configuration spaces.

 Hence for small assemblies, the basic EASAL \cite{Ozkan2011, Ozkan2014MainEasal}
 addresses the demand for a method that
satisfies two criteria:
it should (i) generate a comprehensive roadmap of the assembly configuration space as a topological
complex of constant-potential-energy regions, 
their neighborhood relationships and boundaries; 
and (ii) explicitly formalize and leverage the geometric 
simplicity of these regions (in the case of assembly relative to folding)
to give an efficient and accurate
computation of their volume
by isolation of the region and its boundaries and customized sampling. 
In order to effectively combine the complementary advantages of EASAL with
the abovementioned prevailing methods, the goal of this paper is to 
maintain these advantages of EASAL and 
    Cayley sampling while ensuring 
    certain minimum distance and coverage relationships between sampled
    points in Cartesian space.

\section{Methodology} 
\label{sec:methodology}
The first subsection gives background from \cite{Ozkan2011, Ozkan2014MainEasal}
for the theoretical underpinnings of
EASAL's  key features - geometrization, stratification and convexification 
using Cayley parameters - 
culminating in the concept of an \emph{atlas} of an 
assembly configuration space. 
The second subsection analyzes the issues that arise with a preliminary, straightforward 
use of the Jacobian of the map from Cartesian to Cayley parameters.
The third subsection presents a method of adaptive, optimized choice of 
step-size and direction 
in Cayley sampling that compensates for an ill-conditioned Jacobian.

\subsection{Background: Theory underlying EASAL}
We begin with a description of the input to EASAL.
An \emph{assembly system} 
consisting of the following.
\begin{itemize}
\item 
A collection of 
\emph{rigid molecular components},  
drawn from a small set of \emph{rigid component types} (often just a single type).
Each type is a
is specified as the set of 
positions of \emph{atom-centers}, in a local
coordinate system. In many cases, an \emph{atom-center} 
could be the representation
for the average position of a \emph{collection of atoms in a
residue}.
Note that an assembly \emph{configuration} is 
given by the positions and orientations 
of the entire set of $k$ rigid molecular components in an assembly system, 
relative to one fixed component. Since each rigid molecular component has 6
degrees of freedom, a configuration is a point in $6(k-1)$ dimensional
Euclidean space. The maximum number of atom-centers in any rigid molecular 
component is denoted $n$.
\item
The potential energy is
specified using
\emph{Lennard-Jones}  (which includes  
\emph{Hard-Sphere}) 
\emph{pairwise potential energy functions}. 
The pairwise Lennard-Jones term
for a pair of atoms, $i$ and $j$, 
one from each component, 
is given as a function of the distance 
$d_{i,j}$ between $i$ and $j$; The function is 
typically discretized to take different constant values on 
3 intervals of the distance value $d_{i,j}$:
$(0,l_{i,j}), 
(l_{i,j}, u_{i,j}), and
(u_{i,j}, \infty).$  
Typically, $l_{i,j}$ is the so-called Van der Waal or
steric distance given by "forbidden" regions around atoms 
$i$ and $j.$ And $u_{i,j}$ is a
distance where the 
interaction between the two atoms is no longer 
relevant.
Over these 3 intervals respectively, the Lennard-Jones potential assumes a 
very high value $h_{i,j}$, a small value $s_{i,j}$, and a medium value
$m_{i,j}.$  All of these \emph{bounds} for the intervals for $d_{i,j}$, as well as the values for the 
Lennard-Jones potential on these intervals are \emph{specified constants} as
part of the
input to the assembly model. These constants are specified for 
each pair of atoms $i$ and $j$,
i.e., the subscripts are necessary.
The middle interval is called the \emph{well}.
In the special case of  Hard Spheres, $l_{i,j} = u_{i,j}$. 
\item
A non-pairwise component of the potential energy function in the form of 
\emph{global potential energy} 
terms that
capture other factors 
including the 
implicit solvent 
(water or lipid bilayer
membrane) effect \cite{Lazaridis_Karplus_1999, Lazaridis_2003,
Im_Feig_Brooks_2003}. These are specified 
as a function of the entire assembly configuration.
\end{itemize}
It is important to note that all the above 
potential energy terms are \emph{functions 
of the assembly configuration}. 

\emph{Note} that the input to the
assembly usually specifies the configurations of
interest i.e., a region of the configuration space, 
often specified as a collection $C$ of $m$  atom pairs "of interest"
with the understanding that the only configurations of interest are those in
which at least one
of these $m$ pairs in $C$ occupy their corresponding Lennard-Jones well.
Clearly $m\le n^2\choose{k,2}$.
In addition, we assume the desired level of refinement of sampling is
specified as a desired number of sample configurations $t$.

\subsubsection{Geometrization}  
Observe that for the purposes of this paper stated in Section \ref{intro}, 
it is sufficient to view 
the assembly landscape as a union of constant potential energy regions.
Thus an assembly system can alternatively
be represented as a set of rigid molecular
components drawn from a small set of types, together with 
\emph{assembly constraints}, in the form of distance intervals.
These constraints define \emph{feasible} configurations (where the pairwise
inter-atoms distances are larger than $l_{i,j}$, and any relevant tether and 
implicit solvent constraints are satisfied). The set of feasible
configurations is called the \emph{assembly configuration space}.
The \emph{active constraints} of a configuration are those atom-pairs in the
configuration that lie in the Lennard-Jones well.
An \emph{active constraint} region of the
configuration space is a region consisting of all configurations
where a specified (nonempty) set of constraints is active, i.e, those 
Lennard-Jones 
inter-atom distances between atoms $i$ and $j$ 
lie in their corresponding wells, i.e, the 
interval $(l_{i,j}, u_{i,j})$.

\subsubsection{Stratification, active constraint graphs}  
\label{stratification}
Consider an assembly configuration space $\cal{A}$ of $k$  
rigid components, defined by a system $A$ of assembly
constraints.  The configuration space 
has dimension $6(k-1)$, the number of internal degrees of freedom
of the configurations
since a rigid object in Euclidean 3-space has $6$ rotational and translational
degrees of freedom.
For $k = 2$, this dimension is at most $6$ and 
in the presence of two active constraints, it is at most $4$. 

\begin{figure}
   \centering
   \includegraphics[width=.5\textwidth] {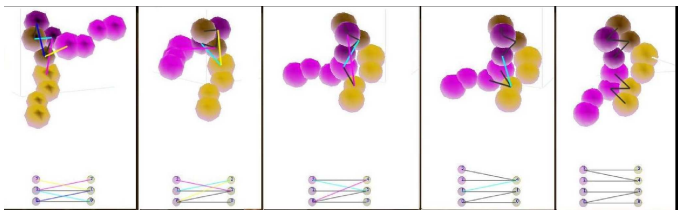}
   \caption{ {\bf Adding constraints, removing parameters until $j=0$}.
   \IT: \Cr s with {\it non-white segments:} parameters 
   and {\it white segements} constraints
   and \IB: activeConstraintGraph \acg\ 
   yielding configurations with ever fewer free parameters 
   as constraints are added one by one.
   }
  \label{fig:contacts}
\end{figure}

A \emph{Thom-Whitney stratification} of the configuration space $\cal{A}$ 
(see \figref{fig:atlas})
is a partition of the space into regions grouped into strata
$X_i$ of $\cal{A}$ that form a filtration 
$\emptyset\subset X_0\subset X_1 \subset \ldots \subset X_m=\cal{A}$,
$m = 6(k-1)$.
%
Each $X_i$ is a union of nonempty closed \emph{active constraint regions}
$R_{Q}$ where 
$m-i$ the set of pairwise constraints $Q\subseteq A$ are {\emph active},
meaning each pair in $Q$ lies in its corresponding Lennard-Jones well, 
and the constraints are independent (i.e., no proper subset of these
constraints generically implies any other constraint in the set).
Each active constraint set $Q$ is itself part of at least one, 
and possibly many, hence $l$-indexed, nested chains of the form
$\emptyset\subset Q^l_0\subset$ $Q^l_1$ $\subset\ldots\subset Q^l_{d-i}=Q$
$\subset\ldots\subset Q^l_m$. 
See Figures \ref{fig:contacts} and \ref{fig:prtree}(left).
These induce corresponding reverse nested chains of active constraint 
regions $R_{Q^l_j}$:
$\emptyset\subset R_{Q^l_d}\subset R_{Q^l_{d-1}} \subset\ldots\subset 
R_{Q^l_{d-i}}=R_Q \subset \ldots\subset R_{Q^l_0}$ 
Note that here for all $l,j$, $R_{Q^l_{d-j}} \subseteq X_{j}$ 
is closed and \emph{effectively} $j$ dimensional; by which we mean that 
if all the $d-j$ Lennard-Jones wells that define the active constraint set 
$Q^l_{d-j}$
narrowed to zero width (i.e, if they degenerated to 
a Hard-Sphere potentials), then the active constraint region
$R_{Q^l_{d-j}}$  would be $j$ dimensional.
%

\def\wid{0.45\linewidth}
\def\wids{0.40\linewidth}
\def\widb{0.55\linewidth}
\begin{figure}
   \centering
   \begin{subfigure}[b]{\wids}
      \epsfig{file=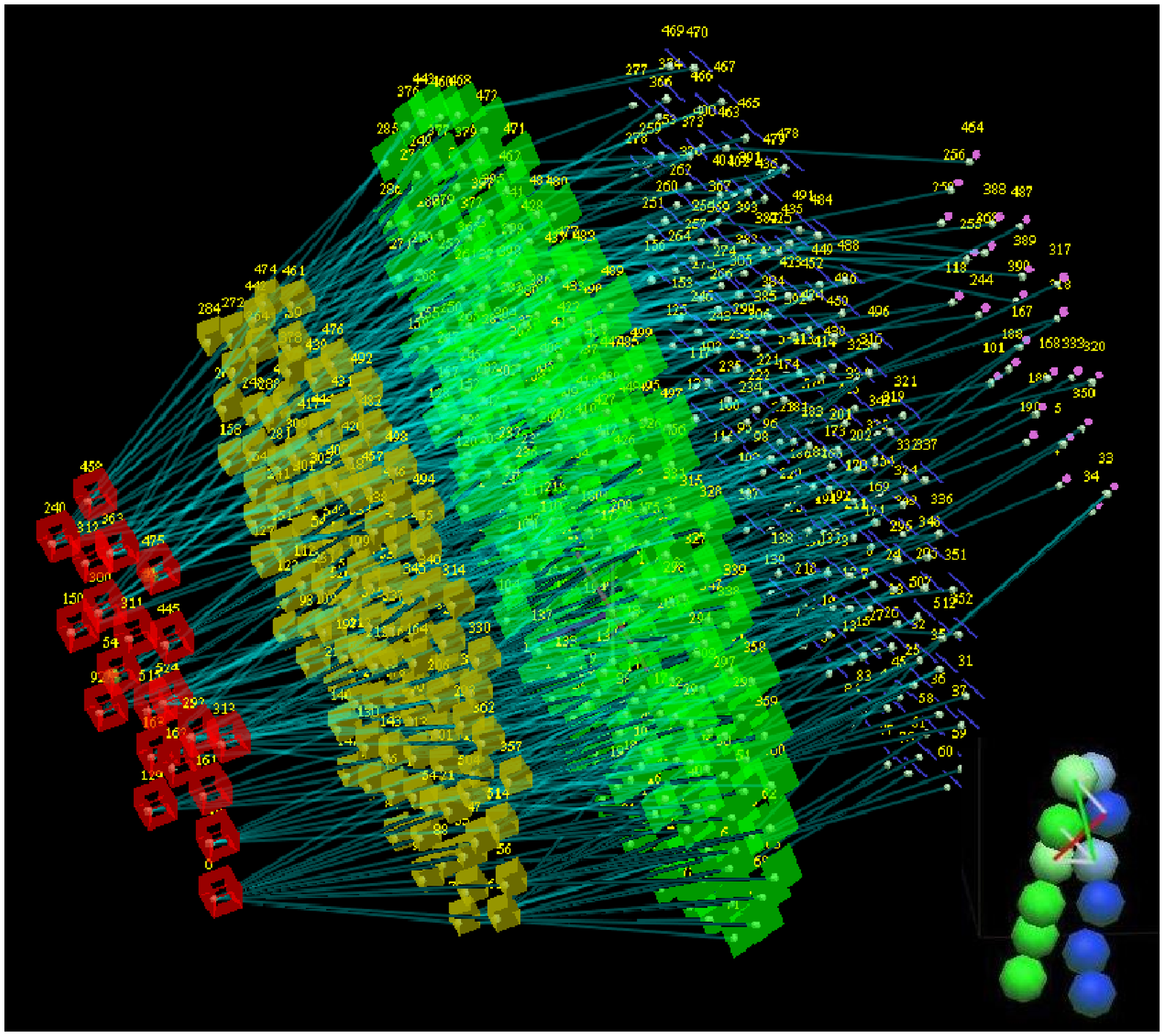, width=\linewidth}
      \caption{stratification of assembly}
      \label{fig:atlas}  
   \end{subfigure}
   \begin{subfigure}[b]{\widb}
      \epsfig{file=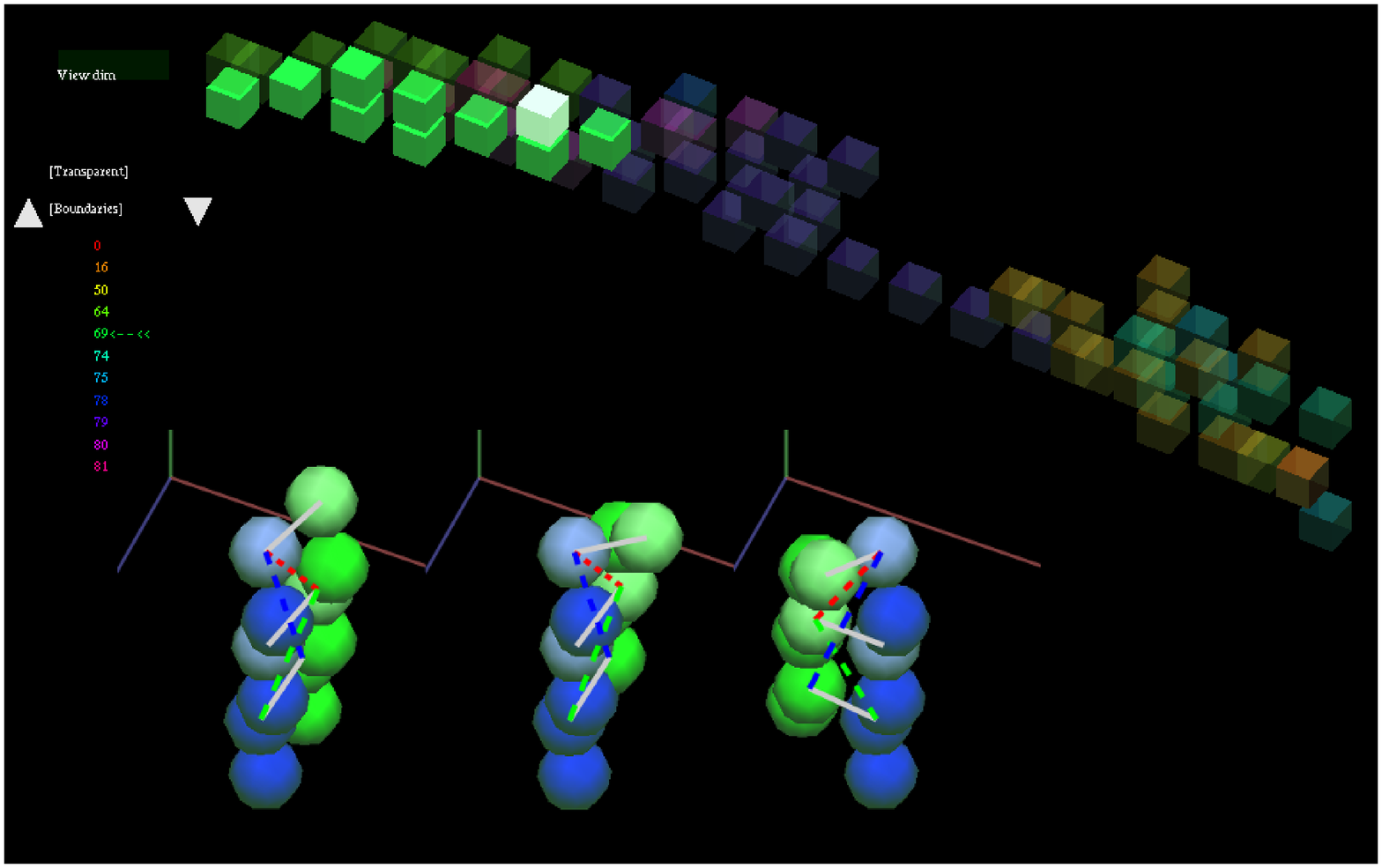, width=\linewidth}
      \caption{ \IT: \Cp s, \IB: \Cr s}
      \label{fig:prtree}
   \end{subfigure}
   \caption{
   (a) {\bf Stratification:} of assembly constraint system
      with parameters $n=$ 
      4 (red), 3 (yellow), 2 (green), 1 (white), 0 (purple). 
   Strata of each dimension $j$ for 
   the assembly constraint system visualized in the lower right inset
   are shown as nodes of one color and shape in a directed acyclic graph.
   Each node represents an active constraint region.
   Edges indicate containment in a parent region one dimension higher.
   (b) {\it top:} Realizable {\bf \Cp s} (distance values) corresponding
   to one node in (a).
   \emph{ Note a different use of color in the display of sample boxes
   in Cayley configuration space than in the stratification diagram.}
   One \Cp\ in the green group is highlighted. 
   {\it bottom:} Three {\bf \Cr s} of the highlighted \Cp.
   Each edge on a realization represents an
   active constraint graph and its chosen parameters.}
\end{figure}

\begin{figure} 
   \epsfig{file=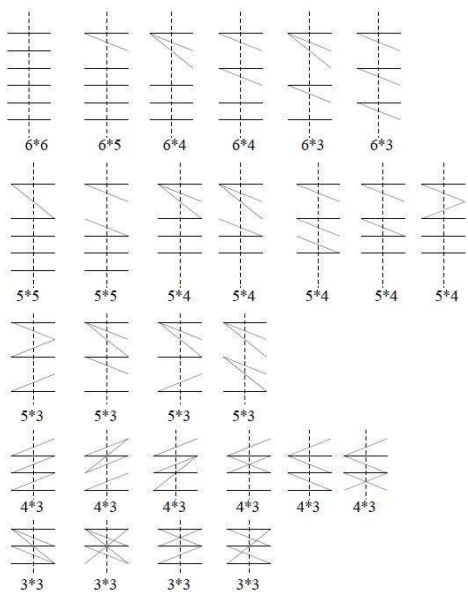, width=3in}
   \caption{All active constraint graphs}
   \label{acgtypes2}
\end{figure}

\begin{figure} 
   \epsfig{file=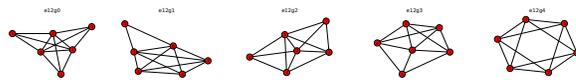, width=3in} 
   \caption{All non-isomorphic active constraint graphs with 6 vertices and 12 edges.}
   \label{fig:v6e12}
\end{figure}

We represent the active constraint system for a region, by an \emph{active
constraint graph} (sometimes called \emph{contact graph}) whose vertices represent the participating atoms 
(at least $3$ in each rigid component) and edges representing the active constraints
between them. Between a pair of rigid components, there are only a small number of 
possible active constraint graph isomorphism types since there are 
at most $12$ contact vertices. For the case of $k=2$ these are listed in Figure
\ref{acgtypes2}, and for higher $k$ a partial list appears in
Figure \ref{fig:v6e12}.

There could be regions of the stratification of 
dimension $j$ whose number of active constraints exceeds 
$6(k-1) -j$, i.e.\ the active constraint system is overconstrained,
or whose active constraints are not all independent. 
Dependent constraints diminish the set of realizations.
For entropy calculations, these regions should be tracked explicitly,
but in the present paper, we do not consider these overconstrained regions 
in the stratification. Our regions are 
obtained by choosing any $6(k-1)-j$ independent active constraints.
%
\subsubsection{Convex representation of active constraint region and atlas}
\begin{figure}
   \centering
   \begin{subfigure}{\wid}
   \epsfig{file=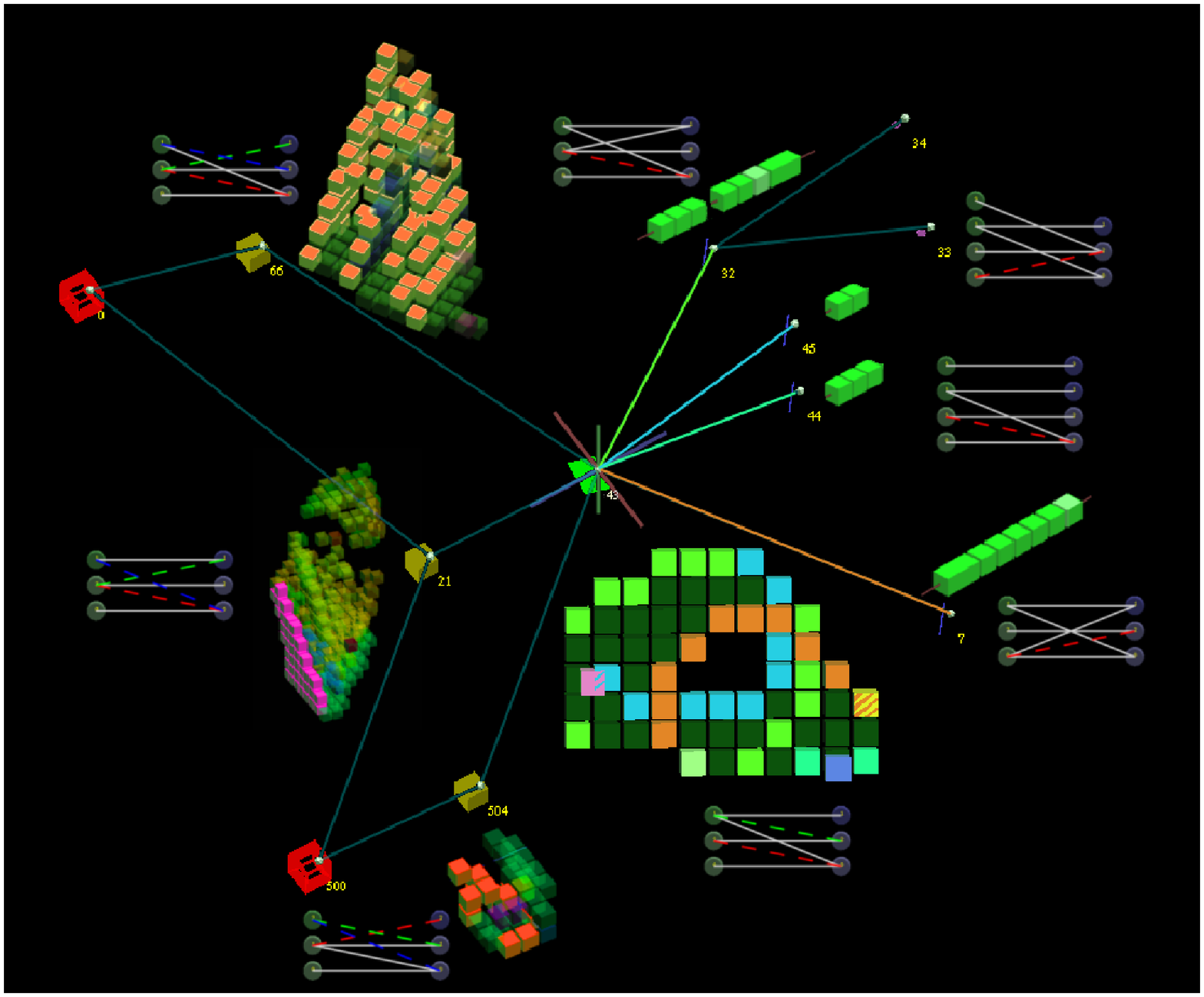, width=\linewidth}
   \caption{
   Cayley \chartt s 
   of dimensions 1,2,3 attached to nodes.}
   \label{fig:pctree}
   \end{subfigure}
   \hskip0.01\linewidth
   \begin{subfigure}{\wid}
   \epsfig{file=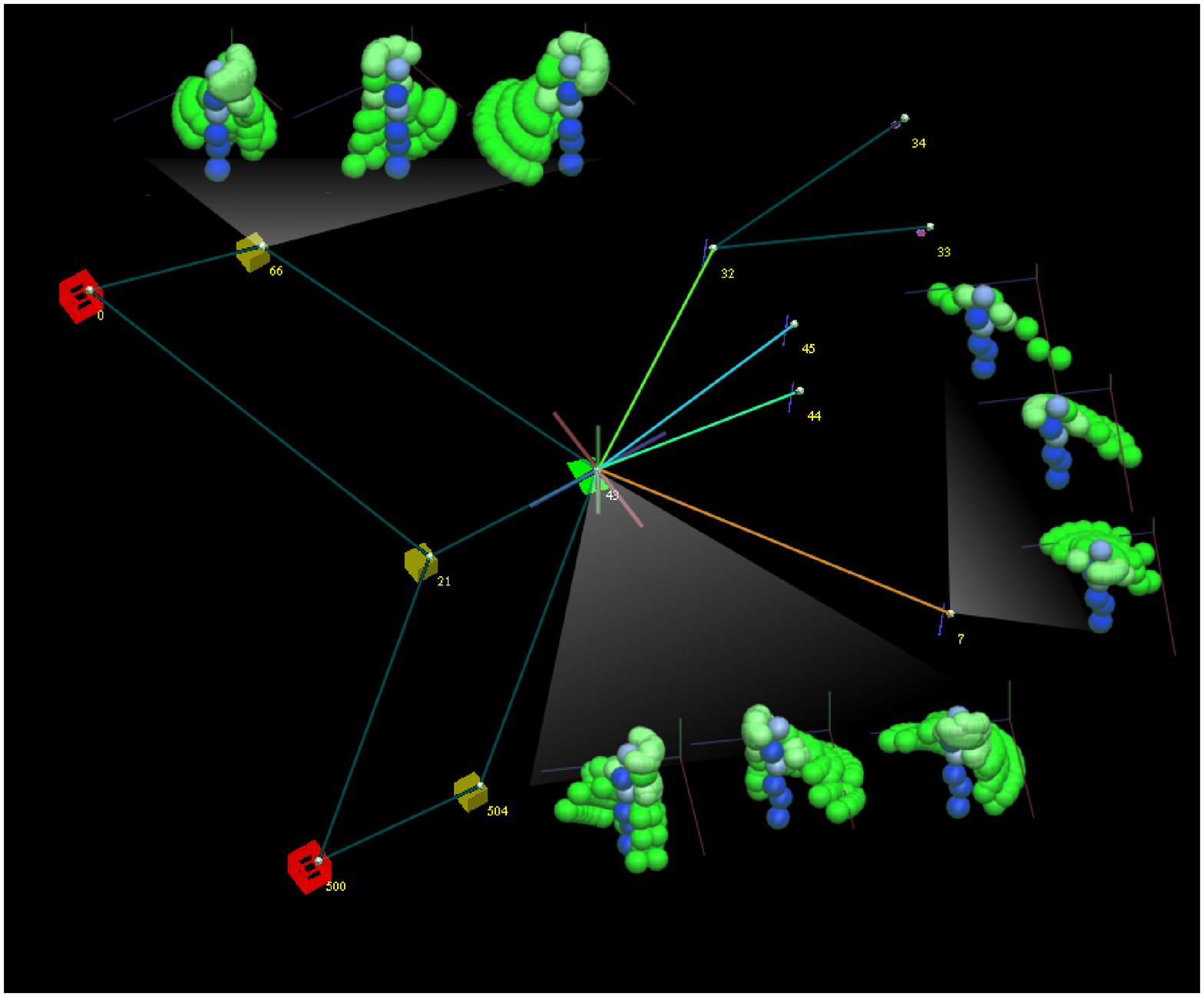, width=\linewidth}
   \caption{\Cr s 
   of dimensions 1,2,3 attached to nodes.}
   \label{fig:flips}
   \end{subfigure}
   \caption{ {\bf Nested chains for one region
   of the \atlas,} i.e.\ nodes and paths in the directed
   acyclic graph of the stratification containing a 2d contraint region.
   {\it center, green:} a $2$d active constraint region.
   {\it left, red and yellow:} 4d and 3d parent regions containing the 2d region.
   {\it right:}  1d and 0d child regions.
   The \acg\ and \chartt\ are displayed next to each region.
   (a)
   The $2$-dimensional (exact, convex) \chartt\ in the center
   has a hole due to infeasible configurations also
   defined by Cayley parameter ranges, hence convex.
   Also, due to choice of different Cayley parameters, the same 
   2-dimensional region appears, without hole, in the $3$-dimensional 
   parent \chartt s as orange boxes {\it top left}, pink boxes {\it middle left}
   and red-orange boxes {\it lower left}; 
   green boxes {\it on right:} 1-dimensional subregions.
   (b)
   Three grey fans attach the \Cr s to their nodes
   as separate sweeps for different chirality of a region
   (the blue \helix\ is fixed without loss of generality). 
   }
\end{figure}

\begin{figure}
\centering
 \begin{subfigure}{3.3in}
\epsfig{file = 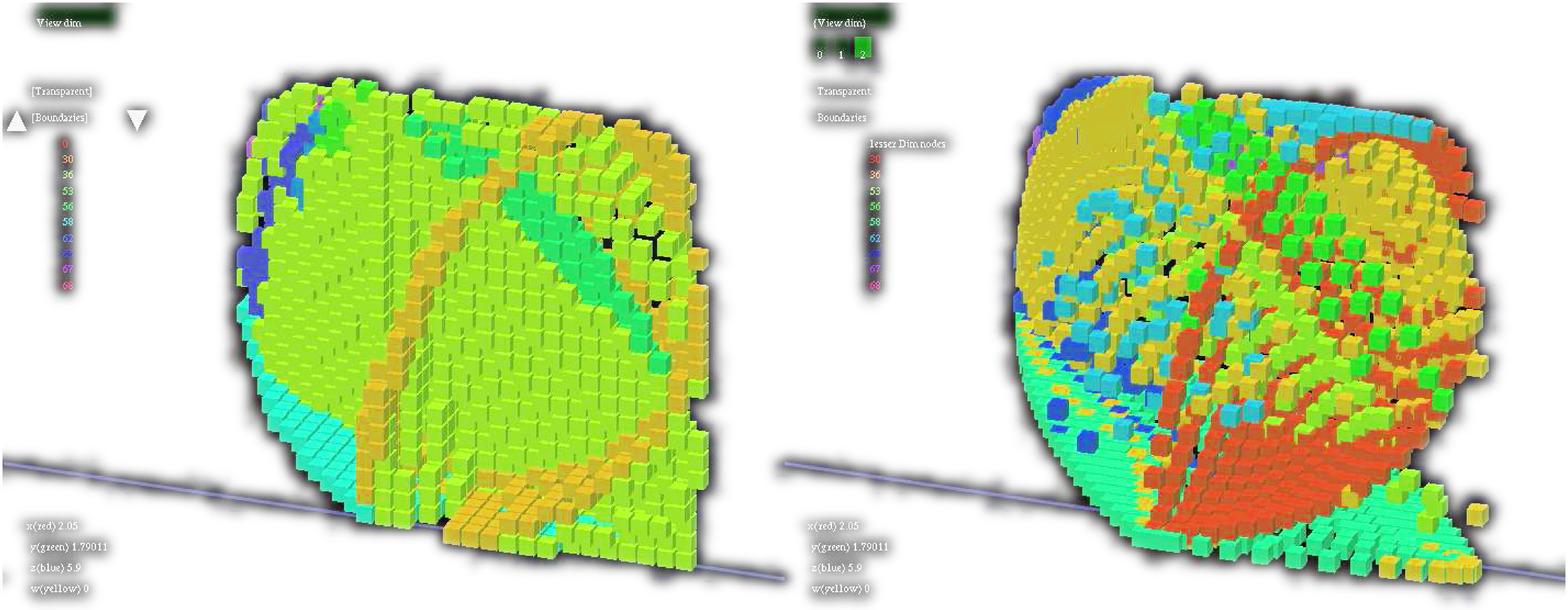, width = \linewidth }
\caption{}
   \label{fig:chart}
   \end{subfigure}
 \hskip0.01\linewidth
\begin{subfigure}{1.4in}
\epsfig{file = 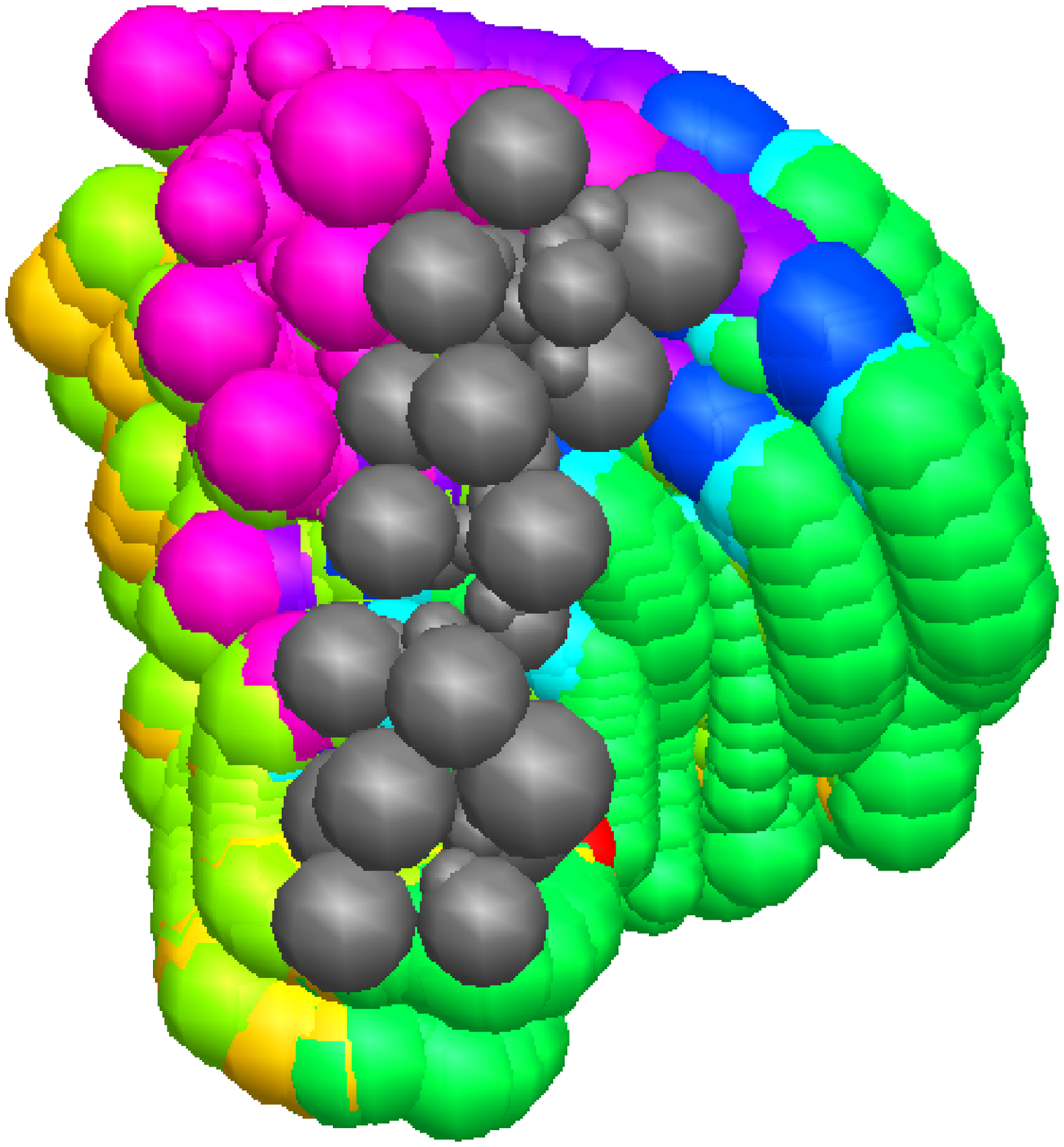, width = \linewidth }
\caption{}
   \end{subfigure}
\begin{subfigure}{1.5in}
\epsfig{file = 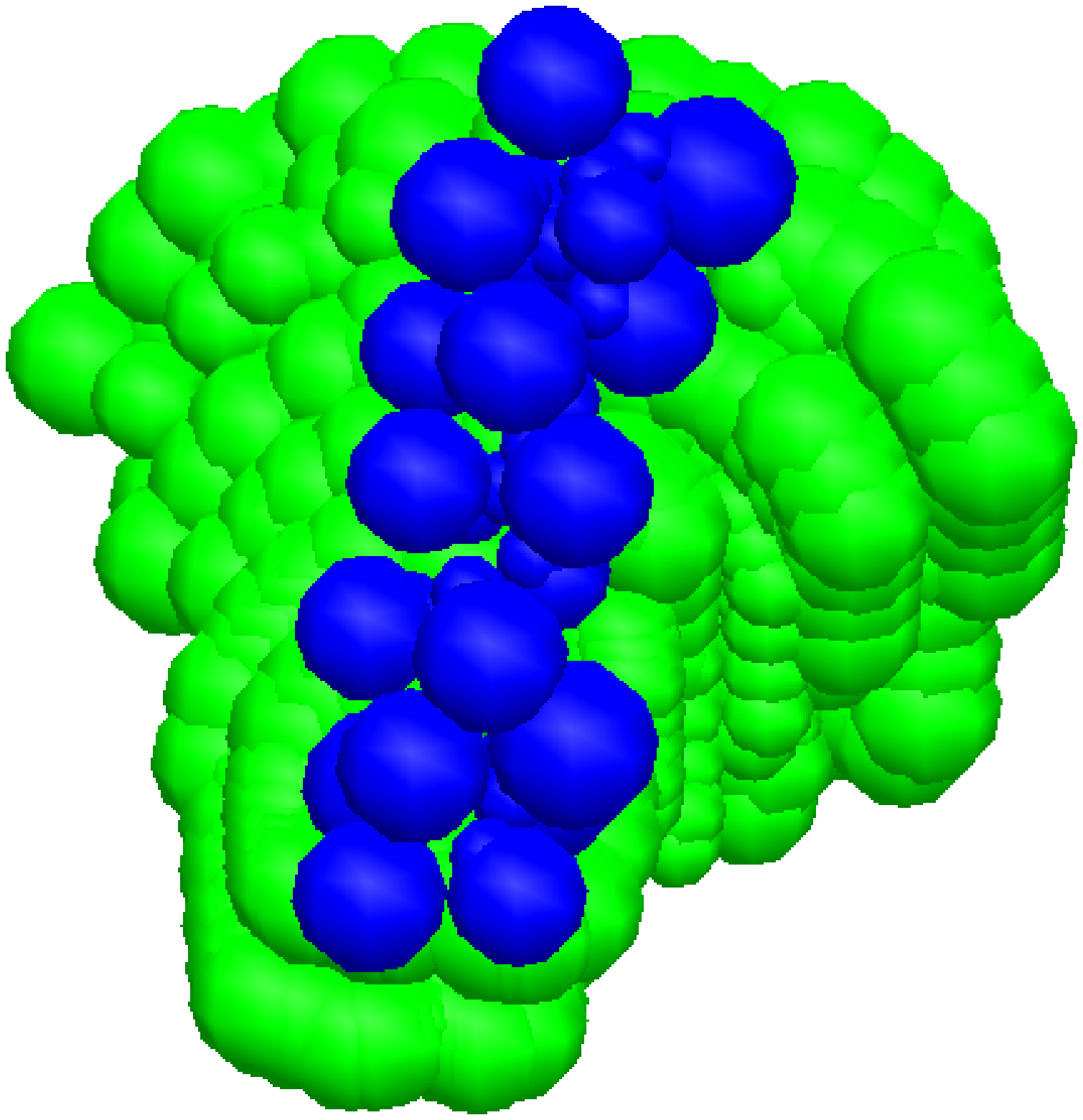, width = \linewidth} 
\caption{}
   \end{subfigure}
\begin{subfigure}{2.25in}
\epsfig{file = 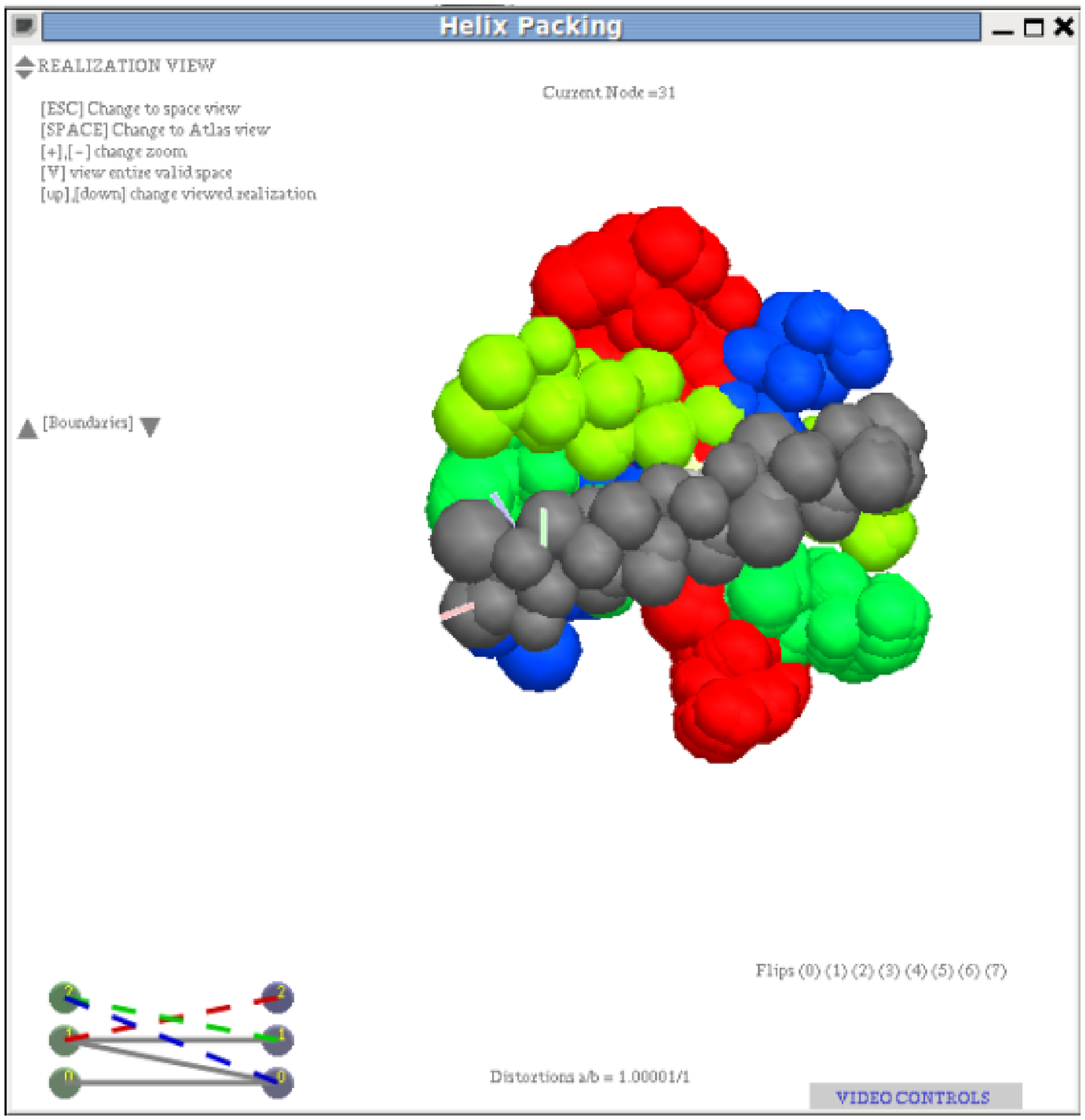, width = \linewidth } 
\caption{}
   \label{fig:reparametrization}
   \end{subfigure}
\caption{\scriptsize Top Left: atlas region showing interiors and boundaries
sampled in its convexifying Cayley parameters; 
boundary/child regions sampled in their own Cayley parameters and mapped
back to the parent region's Cayley parameters ({\sl note increase in samples}). 
Top Right: boundary/child regions sampled in their own Cayley parameters shown as
sweeps around grey reference (toy) helix. 
Bottom Left:
union of boundary regions sampled in parent's Cayley parameters, shown as sweep around blue reference helix ({\sl notice (b) is bigger})
Bottom Right: sweep of one of the boundary 
regions sampled in parent's Cayley parameters is shown in red around gray 
reference helix; the sampling {\sl misses the other
                  colored configurations} in the same boundary region,
                                    obtained by sampling in its own
                                    Cayley parameters.} 
\label{parameterchoice} 
\end{figure}

A new theory of Convex Cayley Configuration Spaces {\bf
(CCCS)}
  recently developed by the author \cite{SiGa:2010}
gives a clean characterization of active constraint graphs 
          whose configuration spaces 
            are convex when represented by a specific choice of
            so-called {\sl Cayley parameters}
              i.e., distance parameters between
                pairs of atoms (vertices in the active constraint graph) that are inactive
                  in the given active constraint region (non-edges in the
                  active constraint graph).
                    See Figure \ref{parameterchoice}. Such active
                    constraint regions are said to be \emph{convexifiable},
                    and the corresponding Cayley parameters are said to be
                    its \emph{convexifying} parameters.
                    See Figures \ref{fig:pctree} \ref{fig:flips}

In general, the active constraint regions $R'_{G}$ 
for an active constraint graph $G$,
can be entirely convexified after ignoring
the remainder of the assembly constraint system,
namely the \aU s not in $G$ and their constraints.
\figref{fig:chart}
The true active constraint region $R_{G}$ is 
subset of $R'_{G}$, however the cut out regions are also defined by active
constraints, hence they, too, could be convexified. 
See Figures \ref{fig:pctree}, \ref{fig:flips}.

When a constraint (edge $e$) not in $G$
becomes active (at a configuration $c$ in $R'_{G}$), 
$G\cup \{e\}$ defines a child active constraint region $R_{G\cup e}$ 
containing $c$. This new region belongs to the stratum 
of the assembly configuration space that is 
of one lower dimension (\defref{stratification}) and defines
within $R'_{G}$ a boundary of the smaller, true active constraint region
$R_{G}$.
%
%
We can still choose the \chartt\ of $R'_{G}$ as 
tight convex \chartt\ for $R_{G}$, but now region $R_{G\cup e}$ has an
exact or tight convex \chartt\ of its own. 
Then the configurations in the region $R_{G\cup e}$ have lower potential
energy since the configurations in that region lie in one more Lennard-Jones
well. Hence they should be carefully sampled in 
free energy and entropy computations although the region 
has one lower effective dimension (e.g, represents a
much narrower boundary channel). However, sampling in the larger parent chart 
of $R(G)$
(of one higher
effective dimension) often does not provide adequate coverage of the 
narrow boundary region 
 $R_{G\cup e}$. 
For example, \figref{fig:reparametrization} shows that
providing a separate \chartt\ for each active constraint region 
can reveal additional realizations at the same level of sampling. 

The {\em Atlas} of an assembly configuration space 
is a stratification of the configuration space 
into convexifiable regions. 
In \cite{Ozkan2011}, we have shown that  
{\sl molecular assembly configuration spaces with 2 rigid
molecular components have an atlas.} 
The software EASAL (Efficient Atlasing and Search of Assembly Landscapes) 
efficiently finds the stratification,  incorporates
provably efficient algorithms to choose the Cayley
parameters \cite{SiGa:2010} that convexify an active constraint region,
efficiently computes bounds for the parametrized convex regions
\cite{ugandhar}, and converts the parametrized configurations into
standard cartesian configurations \cite{eigr2004}.

\subsection{Preliminary Method: Cayley Sampling for Cartesian Uniformity}
We discuss a preliminary method that highlights the issues and challenges
that need to be addressed.

The \Cp s of the \atlas\ need to be converted to \Cr s as in Figure
\ref{fig:flips}.
An assembly configuration is a point in 6 dimensional Cartesian space
representing the rotations and translations of one rigid \helix\ with respect
to the fixed rigid \helix: ($x$, $y$, $z$, $\phi$, $\cos(\theta)$, $\psi$). 
For the active constraint graphs that occur in assembly
\cite{Ozkan2011, Ozkan2014MainEasal}, 
the Cartesian or Euclidean realization can be found using a sequence of
tetrahedra constructions. 

\begin{observation}
    Every \Cp\ in the exact convex \chartt\
    $\Phi_H(G,F,d_F,d_H)$ has at least 1 and generically
    at most finitely many \Cr s in the region
    $R_{G_F}$. See Figure \ref{fig:prtree}.
\end{observation}

Multiple Cartesian orientations correspond to same Cayley configuration. Those orientations are called flips of the Cayley configuration.
The methods to be discussed below executed Cartesian sampling 
on each flip seperately. The flips can meet and bifurcate.
For accurate configurational entropy computations, 
sampling in Cartesian space should maintain a measure of uniformity.
The Cartesian sampling we aim for would be uniform on each flip. 
Ensuring uniformity when the flips are combined is beyond the scope of this
paper.
See fig.~\ref{fig:flips}.\\

\begin{figure}[h!tbp]  
\centering
\def\wid{0.45\linewidth}
 \begin{subfigure}[b]{\wid}
      \epsfig{file = 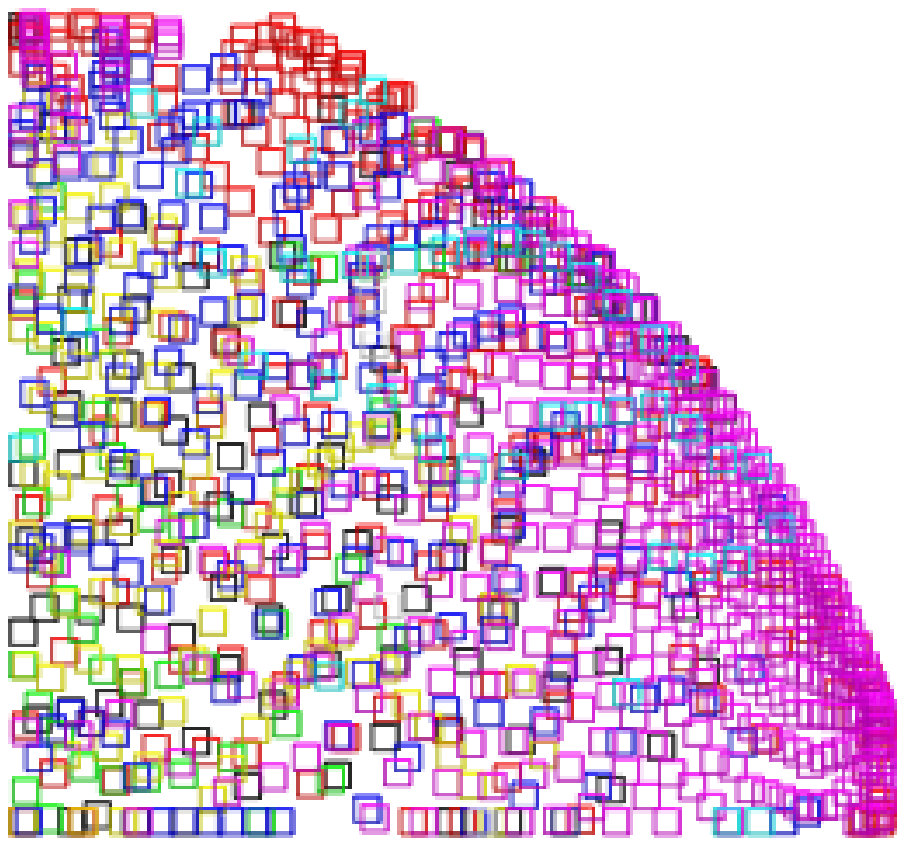, width=\linewidth}
      \caption{$2$-d Cayley Space}
   \end{subfigure}
 \begin{subfigure}[b]{\wid}
      \epsfig{file = 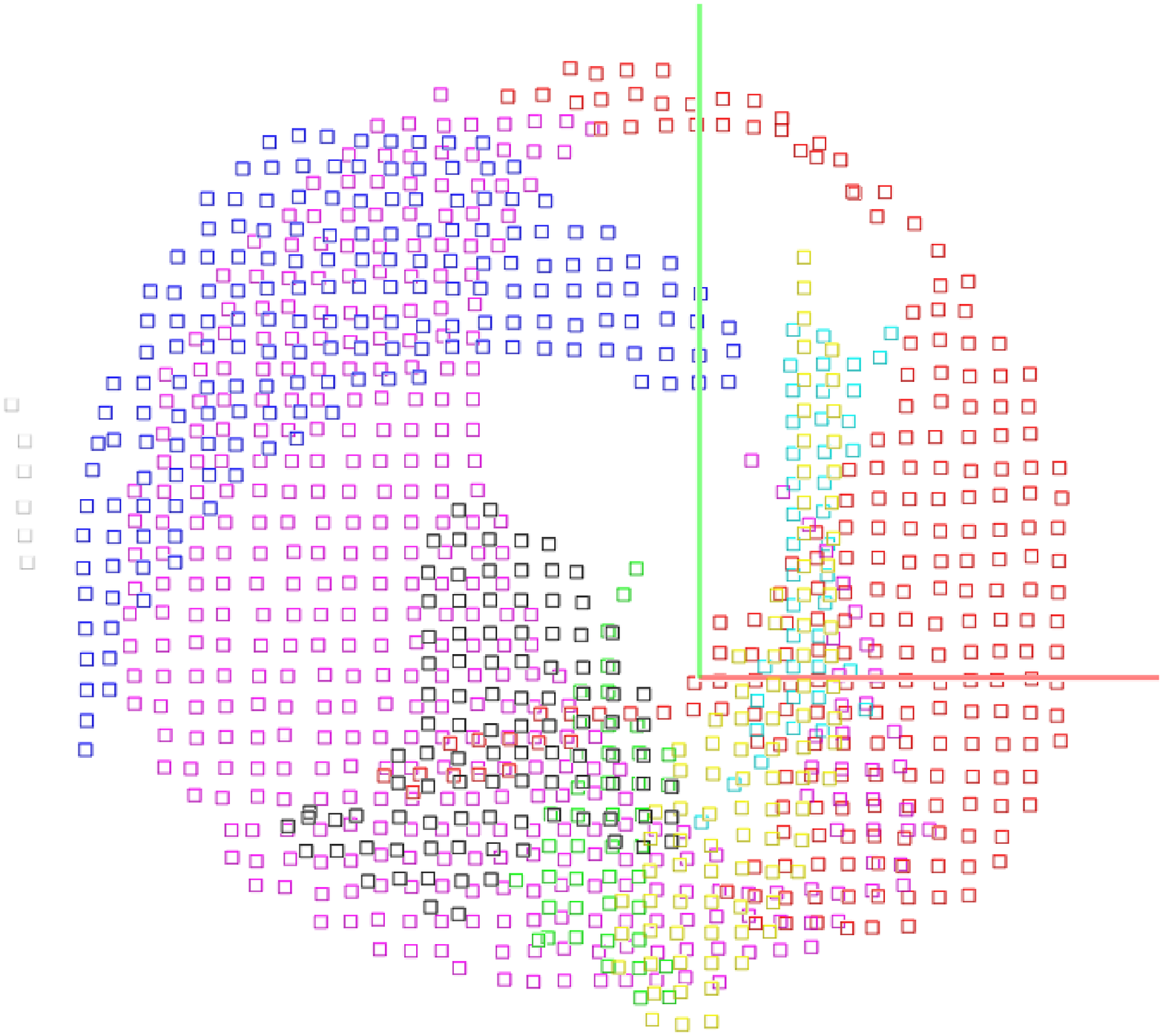, width=\linewidth}
      \caption{Cartesian x, y view}
   \end{subfigure}
%
\caption{ Easal screenshot: a) $2$-D Jacobian sampling projected on
Cayley space. b) $2$-D Jacobian sampling projected on $2$ independent
Cartesian dimensions. c) $2$-D Jacobian sampling projected on $2$ independent Cartesian dimensions plus $1$ dependent dimension.  All flips are colored differently. }
\label{fig:flips}
\end{figure}

However, while ensuring uniform Cartesian sampling on each flip, 
we would like to retain the advantages of Cayley sampling, 
including convexification of the active constraint regions.
To obtain a measure of 
uniform sampling on Cartesian space while Cayley sampling, 
Cayley steps using the (inverse) Jacobian of the map from Cartesian to Cayley.

\begin{definition} [J]
The numerical Jacobian matrix $J$ defines a linear map F: \textit{Cayley space} $\rightarrow$ \textit{Cartesian space}, which is the best linear approximation of the function F near the configuration $p$. Each column of $J$ represents Cartesian changes after walking one step around $p$ = $(p_1, p_2, p_3, p_4, p_5, p_6)$ on Cayley space where $p_i$ is $i$th Cayley parameter. See Table~\ref{table:Jacobian}. 
i.e. the first row of J is the changes along Cartesian $x$ dimension for each Cayley step.\\

\begin{table}[!htbp]
\large
\begin{center}
  \begin{tabular}{  l  c  c  c  c  c  r  } 
\hline
    &     $p_1$           &      $p_2$      &      $p_3$      &      $p_4$      &      $p_5$      &      $p_6$       \\ \hline
$x$   &     $\frac{\Delta x}{\Delta p_1}$     &   $\frac{\Delta x}{\Delta p_2}$    &   $\frac{\Delta x}{\Delta p_3}$    &   $\frac{\Delta x}{\Delta p_4}$    &    $\frac{\Delta x}{\Delta p_5}$   &    $\frac{\Delta x}{\Delta p_6}$    \\ 
$y$   &     $\frac{\Delta y}{\Delta p_1}$     &   $\frac{\Delta y}{\Delta p_2}$    &   $\frac{\Delta y}{\Delta p_3}$    &   $\frac{\Delta y}{\Delta p_4}$    &    $\frac{\Delta y}{\Delta p_5}$   &    $\frac{\Delta y}{\Delta p_6}$    \\ 
$z$   &     $\frac{\Delta z}{\Delta p_1}$     &      .       &      .       &      .       &      .       &      .        \\ 
$\phi$ &   $\frac{\Delta \phi}{\Delta p_1}$     &      .       &      .       &      .       &      .       &      .        \\ 
$\cos(\theta)$ & $\frac{\Delta \cos(\theta)}{\Delta p_1}$&      .       &      .       &      .       &      .       &      .        \\ 
$\psi$ &   $\frac{\Delta \psi}{\Delta p_1}$     &      .       &      .       &      .       &      .       &      .        \\ \hline 
  \end{tabular}
\end{center}
\caption{Jacobian Matrix J}
\label{table:Jacobian}
\end{table}

\end{definition}

It is clear that the numerical Jacobian can be computed at each Cayley point, 
column-wise by finite differences.

%
%

In other words,  let $s_x$, $s_y$, $s_z$, $s_{\phi}$, $s_{\cos(\theta)}$, $s_{\psi}$ be the sizes of the one step for each dimension on Cartesian space. 
Let $\Delta x$, $\Delta y$, $\Delta z$, $\Delta \phi$, $\Delta \cos(\theta)$,
$\Delta \psi$ be the discretized Cartesian differences after one Cayley step. 
Then let $ k_1 = \Delta x/ s_x$, $ k_2 = \Delta y/ s_y$,  
$k_3 = \Delta z/ s_z$,  $k_4 = \Delta \phi/ s_{\phi}$, $k_5 = \Delta \cos(\theta)/ s_{\cos(\theta)}$, $k_6 = \Delta shi/ s_{shi}$ be 
the coordinates of the Cartesian. \\
As criterion of uniformity, we could require the Euclidean 2-norm step distance 
$\|k_1, k_2, k_3, k_4, k_5, k_6\|$ to be 1.\\

In order to achieve the above, we can try
interpolation and binary search over the Cayley step size. 
This works reasonably well if the active constraint region being sampled is
effectively 1-dimensional.
However, for higher dimensions, 
since sampling is usually done one Cayley parameter at a time,
although the Cartesian spacing may be maintained for samples along 
each Cayley line, 
the Cartesian trajectories corresponding to two Cayley lines
may diverge.

\begin{figure*} 
\def\wid{.4\textwidth}
\centering
 \begin{subfigure}[b]{.43\textwidth}
     \includegraphics[width=\linewidth]{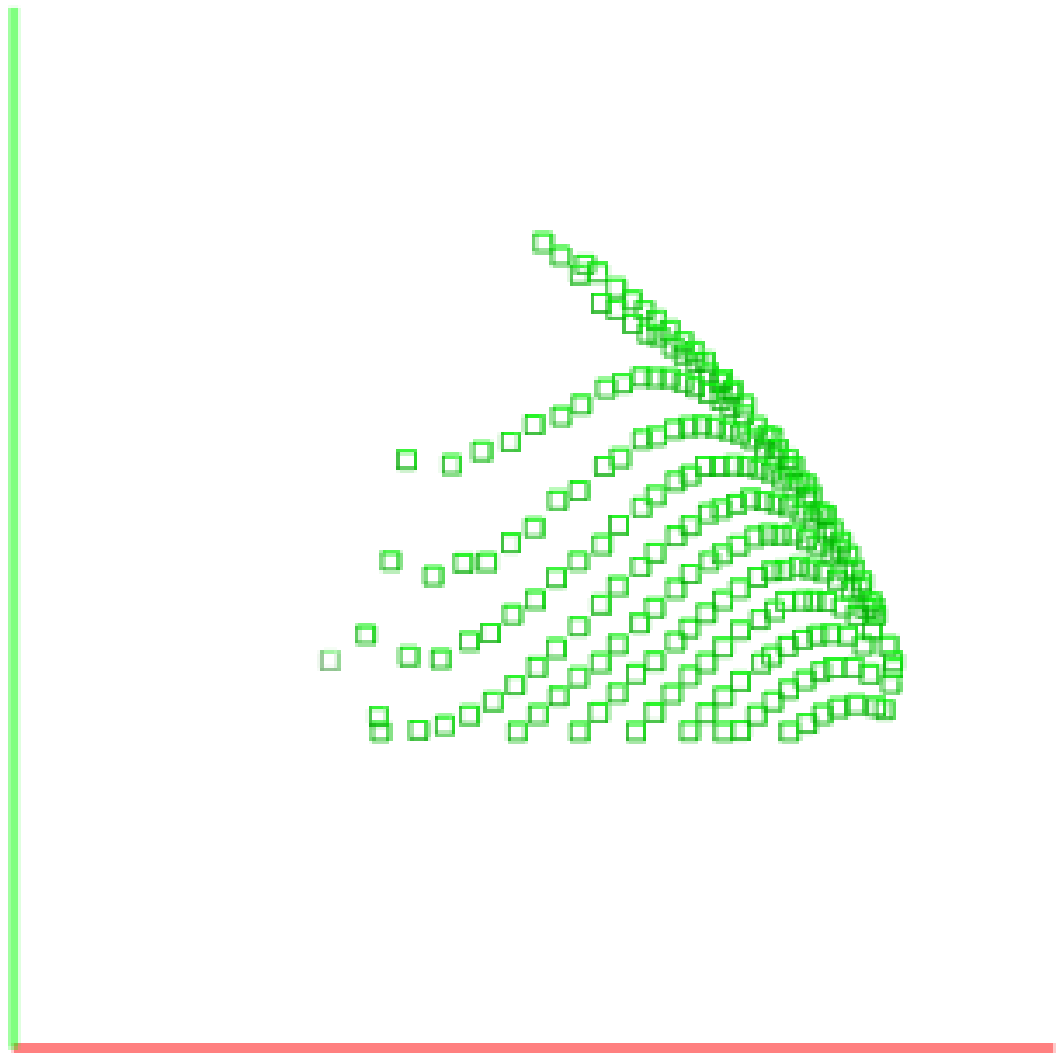} 
      \caption{Uniform Cartesian sampling projected on Cayley Space}
   \end{subfigure}
 \begin{subfigure}[b]{\wid}
      \includegraphics[width=\linewidth]{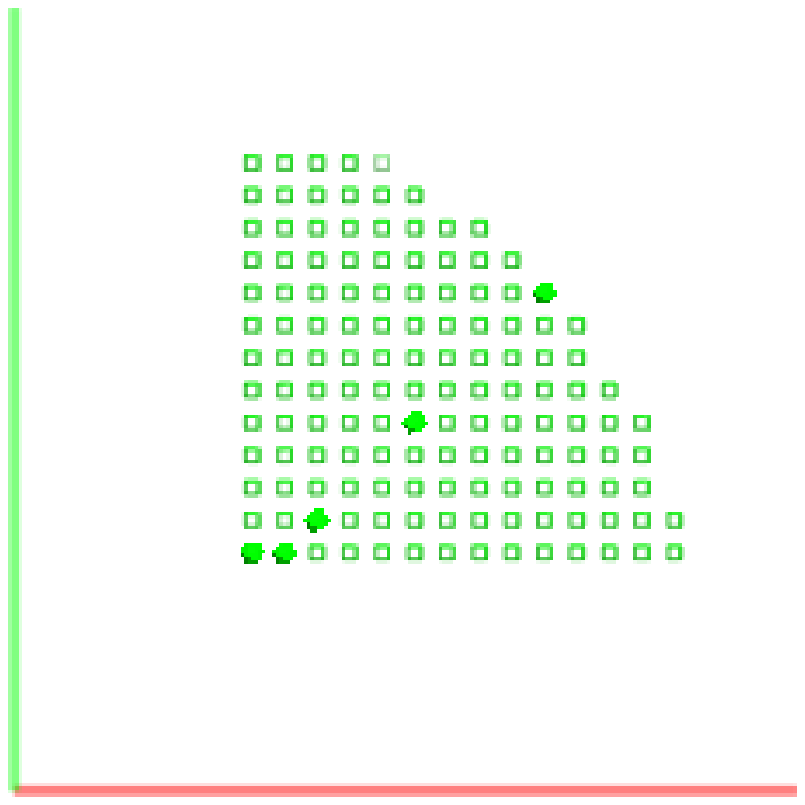} 
      \caption{Uniform Cayley sampling projected on Cayley Space}
   \end{subfigure}
 \begin{subfigure}[b]{\wid}
     \includegraphics[width=\linewidth]{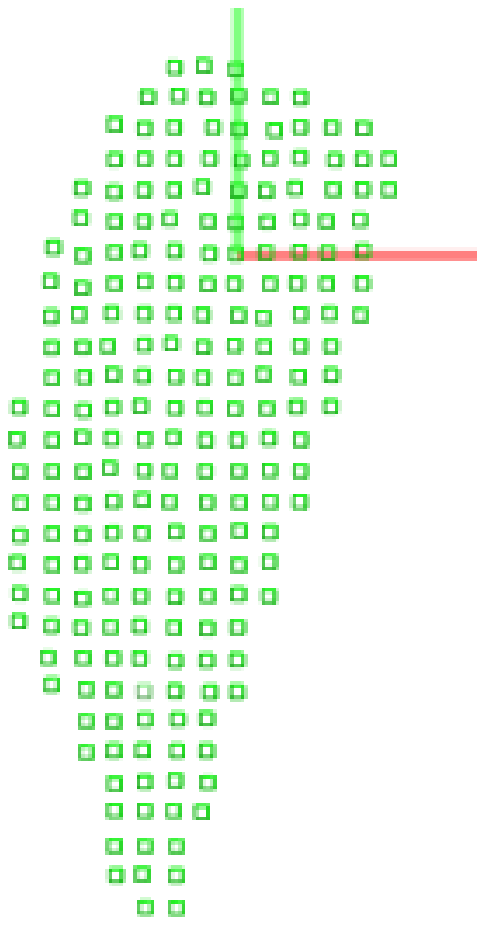} 
      \caption{Uniform Cartesian sampling projected on Cartesian Space}
   \end{subfigure}
 \begin{subfigure}[b]{\wid}
      \includegraphics[width=\linewidth]{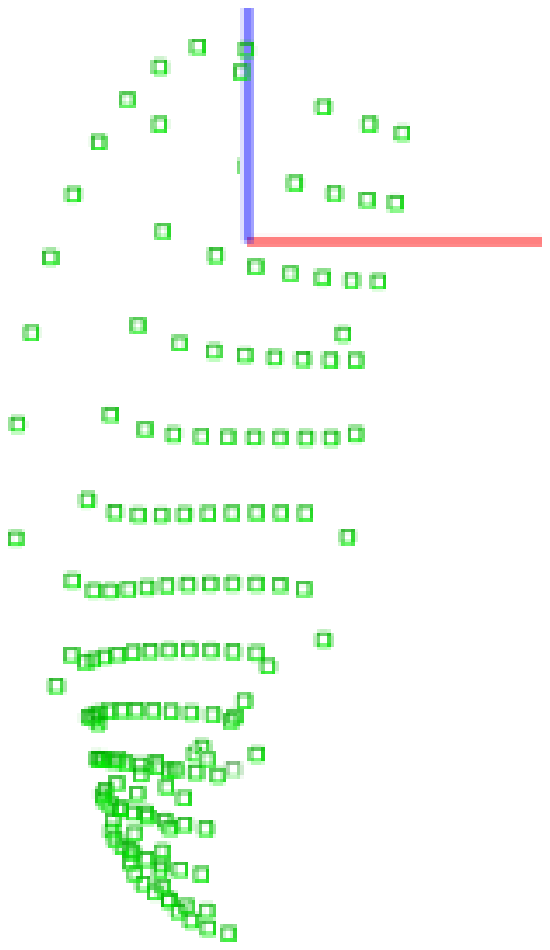}
      \caption{Uniform Cayley sampling projected on Cartesian Space}
   \end{subfigure}
\caption{ Easal screenshot: Different sampling methods projected on both $2$d Cayley space and Cartesian space. Notice the need of walking directionally (not just horizontal and vertical) on Cayley space in order to have uniform sampling on Cartesian space. }
\label{fig:uniformCartesian}
\end{figure*}

In other words, the sampling adjustment should not be restricted only 
to $d$ sampling directions, where $d$ is the effective dimension of 
active constraint region being sampled. 
The entire volume of the $d$-dimensional neighborhood  must be considered
see fig.~\ref{fig:uniformCartesian}, and 
Jacobian adjustments are required to address both the 
step size and direction issues.\\

\begin{definition} 
\textbf{The Orthogonal Cartesian Step Matrix $C$}\\
Let $C$ be the matrix where each column represents 
expected Cartesian changes after one directional Cayley step. 
See Table~\ref{table:CartesianSteps}. 
We would like to walk orthogonally in Cartesian space.\\
\begin{table}[h!tbp]  
\begin{center}
  \begin{tabular}{  l  c  c  c  c  r  } 
	\hline
    $s_x$  &     0    &     0    &     0    &      0      &     0       \\ \hline
     0     &   $s_y$  &     0    &     0    &      0      &     0       \\ 
     0     &     0    &   $s_z$  &     0    &      0      &     0       \\ 
     0     &     0    &     0    &  $s_{\phi}$ &      0      &     0       \\ 
     0     &     0    &     0    &     0    &$s_{\cos(\theta)}$&     0       \\ 
     0     &     0    &     0    &     0    &      0      &   $s_{\psi}$   \\ \hline
 \end{tabular}
\end{center}
\caption{Cartesian Step Matrix $C$: diagonal matrix with Cartesian steps as diagonal entries.}
\label{table:CartesianSteps}
\end{table}
\end{definition}

\begin{definition} 
\textbf{The Cayley Step Matrix $S$ corresponding to $C$}\\
Let $S$ be the matrix of Cayley steps such that when 
adjusted by the Jacobian results in $C$. i.e. $JS = C$. 
$S$ is the numerical $J_{inv}C$. 
See Table~\ref{table:DirectionalCayleyStep}.\\
Each column of $S$ represents one directional Cayley step 
that is predicted to yield orthogonal stepping in Cartesian space.  

\begin{table}[h!tbp]  
\begin{center}
  \begin{tabular}{ l  c  c  c  c  c  r  } 
 \hline
 & $\overrightarrow{s_1}$ & $\overrightarrow{s_2}$ &$\overrightarrow{s_3}$&$\overrightarrow{s_4}$&$\overrightarrow{s_5}$&$\overrightarrow{s_6}$    \\ \hline
$p_1$ & $s_{11}$    &       $s_{21}$    &     .    &     .    &     .    &     .      \\ 
$p_2$ & $s_{12}$    &       $s_{22}$    &     .    &     .    &     .    &     .      \\ 
$p_3$ & $s_{13}$    &       $s_{23}$    &     .    &     .    &     .    &     .      \\ 
$p_4$ & $s_{14}$    &       $s_{24}$    &     .    &     .    &     .    &     .      \\ 
$p_5$ & $s_{15}$    &       $s_{25}$    &     .    &     .    &     .    &     .      \\ 
$p_6$ & $s_{16}$    &       $s_{26}$    &     .    &     .    &     .    &     .      \\ \hline
 \end{tabular}
\end{center}
\caption{Directional Cayley Steps}
\label{table:DirectionalCayleyStep}
\end{table}
\end{definition}

%
%
%
%


\subsubsection{Issues}
\label{issues}

\underbar{Ill-conditioned Jacobian:} \\ 
Jacobian matrix is by definition an \textit{linear approximation} of the 
nonlinear map $F: \textit{Cayley space} \rightarrow \textit{Cartesian
space}$.
The Jacobian can be ill-conditioned and sensitive to small changes and
numerical errors in its arguments.\\

\begin{figure}[h!tbp]
\def\wid{.4\textwidth}
\centering
\epsfig{file = 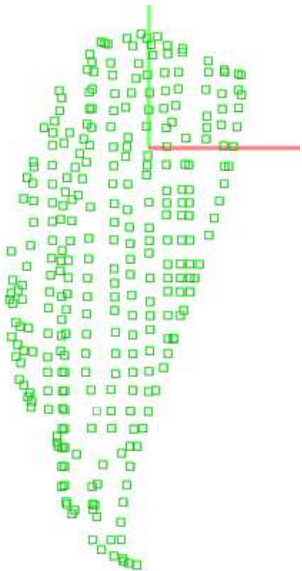, width=.25\textwidth}
\caption{ Easal screenshot: Jacobian sampling projected on Cartesian
Space fails to satisfy uniformity for some regions.}
\label{fig:corrupted}
\end{figure}

\underbar{What Cayley trajectory to follow to ensure}
\underbar{comprehensive coverage?} \\
In uniform Cayley sampling, the Cayley parameters are walked one by one (grid sampling on Cayley space). 
With the above Jacobian adjustments to Cayley step direction, 
such grid sampling is impossible. 
Hence 
it is important to have a systematic method to determine 
what path to follow avoiding repetitions
and ensuring coverage. 
For a single Cartesian dimension the corresponding Cayley direction 
is specified by the Jacobian adjustment in every step. 
As in the previous approach (without direction adjustment)
it is not clear how to generalize this to higher dimensional regions.
While uniform Cayley sampling comprehensively covers 
Cayley space and thereby also Cartesian space.  
This property is not generally preserved 
by the use of Jacobian adjustments to stepping direction.


\subsection{Recursive, Adaptive Cayley Sampling} 
\label{sec:whatdirection}
We propose an Iterative Jacobian computation method with
adapted step magnitude and direction, followed by a recursive Cayley trajectory
determination method to deal with the issues discussed in
the previous subsection.
We will use $S, C, J$  to denote the $d\times d$ matrices of Cayley steps,
Cartesian steps, Jacobian, respectively, as
described above, where $d$ is dimension of the active constraint
region that is currently being sampled. 
Recall that the value $d$ is at most 6 for packing of $2$ rigid molecules.
The first two subsections deal separately with the two issues mentioned above:
Illconditioning of the Jacobian and Cayley sampling trajectory.

\subsubsection{Ill-conditioning: Iterative Jacobian computation}
The Jacobian matrix would give the best approximation, if the Cayley steps that
are used to create Jacobian matrix are close to the output Cayley steps
as a result of Jacobian adjustment. 
In order to achieve best approximation, we iterate on 
Cayley directions and magnitudes until convergence.
See Algorithm~\ref{alg:computeCayleyDirectionIteratively}. \\

\begin{algorithm}[h!tbp]  %
\SetKwInOut{Input}{input}\SetKwInOut{Output}{output}
{\bf computeCayleyDirectionIteratively}\\
  \Input{$S, C$, config   }
  \Output{$S$ \tcp*[f] final Cayley steps }
  \BlankLine
  $J$ := computeJacobian($S$, config) \\
  $K := J_{inv}C$;  \tcp*[f]{$K$ is Cayley transformer(definition below)}\\  
  $S := SK$; \\
  \eIf 
  {$K$ is not close to identity matrix}
  {
  return computeCayleyDirectionIteratively($S, C$, config);
  }
  {return $S$;} 

  \caption{computeCayleyDirectionIteratively} 
\label{alg:computeCayleyDirectionIteratively} 
\end{algorithm}


Note that when the the numerical Jacobian is computed, 
the $i$th column of $J$ represents
Cartesian changes after walking one step on Cayley parameter $p_i$. The 
$i$th column of $J$ is divided to $\Delta p_i$ which is a scalar value. See Table~\ref{table:Jacobian}. 
However, now the $i$th column of $J$ represents Cartesian changes after walking 
one directional Cayley step $\overrightarrow{s_i}$ that is 
$i$th column of $S.$ Hence $\Delta \overrightarrow{s_i}$ is a vector having
components in all Cayley parameters $p_i$. 
So we redefine the Jacobian matrix as:
     
\begin{table}[h!tbp]  
\begin{center}
  \begin{tabular}{  l  c  c  c  c  c  r  }  
\hline
    & $\overrightarrow{s_1}$ & $\overrightarrow{s_2}$ & $\overrightarrow{s_3}$ & $\overrightarrow{s_4}$ & $\overrightarrow{s_5}$ & $\overrightarrow{s_6}$ \\ \hline
$x$   &     $\Delta x_{s_1}$     &   $\Delta x_{s_2}$    &   $\Delta x_{s_3}$    &   $\Delta x_{s_4}$    &    $\Delta x_{s_5}$   &   $\Delta x_{s_6}$    \\ 
$y$   &     $\Delta y_{s_1}$     &   $\Delta y_{s_2}$    &   $\Delta y_{s_3}$    &   $\Delta y_{s_4}$    &    $\Delta y_{s_5}$   &   $\Delta y_{s_6}$    \\ 
$z$   &     $\Delta z_{s_1}$     &      .       &      .       &      .       &      .       &      .        \\ 
$\phi$ &     $\Delta \phi_{s_1}$     &      .       &      .       &      .       &      .       &      .        \\ 
$\cos(\theta)$ & $\Delta \cos(\theta)_{s_1}$  &      .       &      .       &      .       &      .       &      .        \\ 
$\psi$ &   $\Delta \psi_{s_1}$     &      .       &      .       &      .       &      .       &      .        \\ \hline 
  \end{tabular}
\end{center}
\caption{Redefined Jacobian Matrix J}
\label{table:newJacobian}
\end{table}

%
%

\begin{figure}[h!tbp]  
\def\wid{.4\textwidth}
\centering
 \begin{subfigure}[b]{.44\textwidth}
      \epsfig{file = 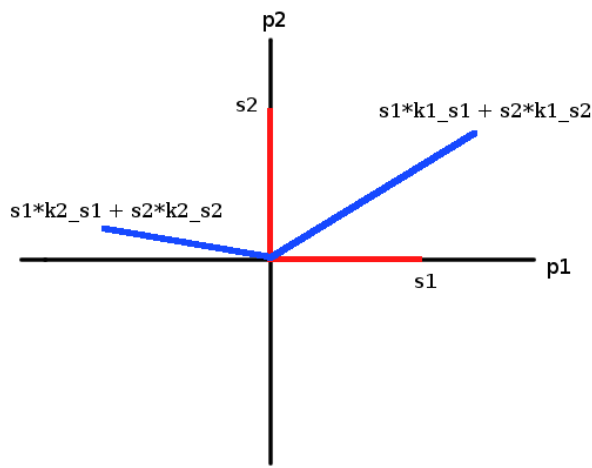, width=\linewidth}
      \caption{$2$D Cayley basis transformation}
   \end{subfigure}
 \begin{subfigure}[b]{\wid}
      \epsfig{file = 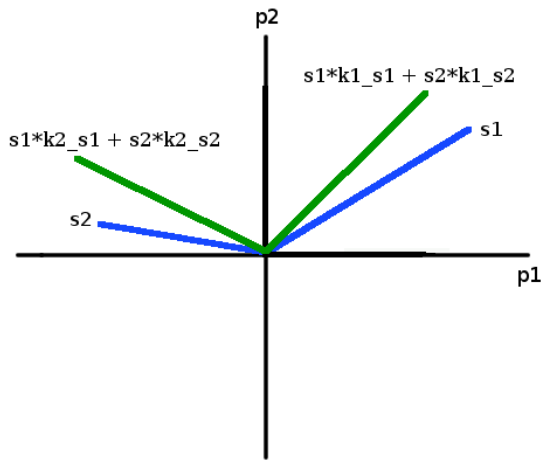, width=\linewidth}
      \caption{$2$D Cayley basis transformation second iteration}
   \end{subfigure}
\caption{ Initial Cayley steps are orthogonal to Cayley parameter
space ($p_1$, $p_2$). $K$ is first applied to inital Cayley steps(red lines) to achieve new Cayley basis(blue lines). Then applied to current Cayley basis (blue lines) to achieve new Cayley basis (green lines). }
\label{fig:CayleyTransformation}
\end{figure}

With the redefined Jacobian $J$ 
$J_{inv}C$ has a new interpretation.
\begin{definition} [The Cayley transformer matrix $K$]
Let $K$ be the Cayley transformer matrix such that when adjusted by the 
Jacobian we obtain $C$. i.e. $JK = C$ \\
See Table~\ref{table:CayleyConverter}.\\
Each column of $K$ contains the coefficients of current Cayley steps that
will lead to new direction in Cayley space that will yield 
orthogonal sampling in Cartesian space. See fig.~\ref{fig:CayleyTransformation}.\\

\begin{table}[h!tbp]  
\begin{center}
  \begin{tabular}{  l  c  c  c  c  r  } 
	\hline
  $k_1\_s_1$    &       $k_2\_s_1$    &     .    &     .    &     .    &     .      \\ \hline
  $k_1\_s_2$    &       $k_2\_s_2$    &     .    &     .    &     .    &     .      \\ 
  $k_1\_s_3$    &       $k_2\_s_3$    &     .    &     .    &     .    &     .      \\ 
  $k_1\_s_4$    &       $k_2\_s_4$    &     .    &     .    &     .    &     .      \\ 
  $k_1\_s_5$    &       $k_2\_s_5$    &     .    &     .    &     .    &     .      \\ 
  $k_1\_s_6$    &       $k_2\_s_6$    &     .    &     .    &     .    &     .      \\ \hline
 \end{tabular}
\end{center}
\caption{Cayley Transformer $K$}
\label{table:CayleyConverter}
\end{table}
\end{definition} 

In order to compute $K$, $J_{inv}$ needs to be computed, hence $J$ has to be a square matrix.
At first glance, computing the inverse of Jacobian matrix can be worrying
since the Jacobian matrix is $6 \times d$ matrix now.
However, if Cayley space is $d$ dimensional ($d < 6$), then in fact 
the Cartesian basis has only $d$ independent vectors
Hence, we can crop $6 - d$ rows of Jacobian matrix to make 
it $d\times d$ square matrix. 
Here the question is then how to best find those dependent $6 - d$ rows.
Among all ${6 \choose d}$ combinations of $d\times d$ submatrix of $J$, 
pick the one that gives best determinant. 

%
%
%
%

Figure~\ref{fig:CayleyTransformation} illustrates the transformation of
from initial orthogonal Cayley basis to the new directed Cayley
basis. At each iteration, new Cayley transformer matrix $K$ is computed.\\

The following method can be used to speed up convergence of the above method 
or for finer adjustments - its convergence, however, is not guaranteed.
It works best for a small
number of dimensions.

\subsubsection{Illconditioning: Adaptive magnitude and direction:}

In order to correct the direction distortions,
the idea is to precompute, for the $i$th direction Cartesian step, 
how much distortion is caused in the $j$th direction. 
Adjust the $j$th direction by using $j$th Cayley step 
that is dedicated to the $j$th Cartesian dimension and 
subtract those distortion adjustments 
\comm{from the $i$th direction Cartesian} step. 
See Algorithm~\ref{alg:adaptiveMagnitudeAndDirection}. \\

\begin{algorithm*}[h!tbp]  %
\SetKwInOut{Input}{input}\SetKwInOut{Output}{output}
{\bf adaptiveMagnitudeAndDirection}\\
  \Input{$S$, $C$, config, $i$, threshold \tcp*[f]{$i$:direction} }
  \Output{ out\_config  }
  \BlankLine
	new\_config := adaptiveMagnitude($S(i), C(i,i)$, config, $i$, threshold); \\
	\If{ new\_config is still failed due to inaccuracy of Jacobian}
        { return failed}
	$\gamma$ := the distortions on all Cartesian directions between new\_config and expected new\_config \\
	$\gamma_c$ :=  $\gamma$ corresponding to the deviation in terms Cartesian unit steps. \\
	\For{ each $j$th Cartesian direction }
	{
	      set Cayley step $s_j$ to -$S(j)\gamma_c(j)$ to reset the deviation on $j$th direction of Cartesian space. \\
	      temp\_config := adaptiveMagnitude($s_j$, $\gamma(j)$, new\_config, $j$, threshold);\\
	      \If{ able to fix distortion } 
              { update new\_config to be temp\_config consecutively }
	}

	// final check if distorted or not after cumulative distortion fixes \\
	compute the change on Cartesian direction $i$ between $config$ and new\_config \\
	ratio := the change / expected Cartesian step $C(i,i)$ \\
	\eIf{  ratio is within [1 $\pm$ threshold] }
	{   return new config;  }
	{   return failed; }

  \caption{adaptiveMagnitudeAndDirection} 
\label{alg:adaptiveMagnitudeAndDirection} 
\end{algorithm*}

The method Algorithm~\ref{alg:adaptiveMagnitude} called above 
has \textit{adaptive step size} to compensate for the inaccuracy of the Jacobian. 
It uses binary search on the step size (multiplier to the column) until it gets the desired step size.
The adaptive search stops if stepping ratio is within [1 $\pm$ threshold]. \\

\begin{algorithm*}[h!tbp]  
\SetKwInOut{Input}{input}\SetKwInOut{Output}{output}
{\bf adaptiveMagnitude}\\
  \Input{$s_i$, $c_i$, $config$, $i$, $threshold$, $min\_s_i$, $max\_s_i$ \tcp*[f] {$s_i$: $i$th Cayley step, $c_i$: $i$th Cartesian step, $i$:direction} }
  \Output{ out\_config  }
  \BlankLine
  update $min\_s_i$ and $max\_s_i$ by comparing $s_i$\\
  \If{ $min\_s_i$ $>$ $max\_s_i$  }
  { return failed;}

  compute new\_config by walking from $config$ by step := $s_i$ \\
  \eIf{ new\_config is realizable }
  {
	  compute the change on Cartesian dimension $i$ between $config$ and new\_config \\
	  ratio := the change / expected Cartesian step $c_i$ \\
	  \eIf 
	  {   ratio is within [1 $\pm$ threshold]  }
	  {   return new config;  }
	  {   $s_i$ := $s_i$ / ratio;\\ 
	      return adaptiveMagnitude($s_i$, $c_i$, $config$, $i$, $threshold$, $min\_s_i$, $max\_s_i$); 
	  }
  }
  {
	$s_i$ := ($min\_s_i$ + $max\_s_i$)/2 ;\\ 
	return adaptiveMagnitude($s_i$, $c_i$, $config$, $i$, $threshold$, $min\_s_i$, $max\_s_i$); 
  }
  \caption{adaptiveMagnitude} 
\label{alg:adaptiveMagnitude} 
\end{algorithm*}

As mentioned earlier, these patch ups work well in practice 
for fine tuning or for 
small number of dimensions. Convergence is not guaranteed in theory. 
For the high dimensions, the adjustment of one dimension may increse
distortion in another. \\
Hence, correctness of input Cayley direction matrix $S$ is crucial,
for which the Cayley trajectory becomes important, in order to 
achieve the best approximation of $S$ 
by recursive Jacobian computation in
Algorithm~\ref{alg:computeCayleyDirectionIteratively} above.\\
We discuss the issue of Cayley trajectory next.

\subsubsection{Recursive Cayley trajectory}
\label{sec:whatpath}
Recursive Cayley sampling walks in every direction of 
Cartesian space at every step. 
If it hits a boundary, then it does not proceed forward at that point.
Since in our assembly settings, feasible regions in 
Cartesian space are connected 
RecursiveSampling will find a path to cover the region.\\
This way, in the case of a nested infeasible region inside a feasible region 
such as a steric boundary, just the boundary of the infeasible region 
is sampled (the inside of the steric region is not inefficiently sampled and 
discarded).\\ 

In order to keep track if specific points in Cartesian grid have been 
visited, a boolean map $M$ is used as Cartesian grid coordinate system of
appropriate size. 
See Algorithm~\ref{alg:RecursiveSampling}. \\


\begin{algorithm*}[h!tbp]  %
\SetKwInOut{Input}{input}\SetKwInOut{Output}{output}
{\bf RecursiveSampling}\\
  \Input{$S$, $C$, $M$, cartpoint, config, threshold \tcp*[f]  
  {$M$:the matrix keeps track of a point is visited or not, cartpoint: the
  vector for current Cartesian point in $M$} }
  \Output{$R$ \tcp*[f] set of configs }
  \BlankLine
	$R$ := $R$ + config; \\	
	$new\_S$ := computeCayleyDirectionRecursively($S$, $C$, config); \tcp*[f] uses S from previous point to converge faster. \\
	\For{each Cartesian dimension $i$ and reverse direction of $i$}
	{
		new\_cartpoint:= cartpoint; \\
		new\_cartpoint($i$) := $\pm$1; \tcp*[f] -1 if reverse direction \\
		\If { $M$(new\_cartpoint) is not visited before }
		{
			new\_config := adaptiveMagnitudeAndDirection($new\_S$, $C$, config, $i$, threshold); \\
			\If{ new\_config is failed \tcp*[f] due to Cartesian boundary or Jacobian was not good enough approximate to walk one step within the threshold}
			{	
				new\_config := jumpToDisconnectedRegion($S(i)$, $M$, cartpoint, config ); \tcp*[f] uses $S(i)$ that is previous Cayley step \\
				update new\_cartpoint of new\_config; 
			}
			\If{ new\_config is succesfully computed}
			{
				set M(new\_cartpoint) to true to be visited \\	
				\If { new\_config doesnot hit any boundary like sterics or etc.}
				{ 	
					RecursiveSampling($new\_S$, $C$,
                                        $M$, new\_cartpoint, new\_config, threshold);  \\
				}
			}
			
		}
	}

  \caption{RecursiveSampling} 
\label{alg:RecursiveSampling} 
\end{algorithm*}

\textbf{Note:}
Consecutive small deviations that are within a tolerance 
at each step may result 
in change of the direction of the path.
In order to correct this:\\
Usually, expected step size is set to be $C(i,i)$ for $i$th Cartesian
direction in Algorithm~\ref{alg:adaptiveMagnitudeAndDirection}.
However if previous point is deviated for the amount of $\mu$ from the
original path along an arbitrary Cartesian direction, then the next step
size should be set to $C(i,i) - \mu$. \\

\medskip\noindent
\underbar{Narrow Cartesian Gates}
As pointed out earlier, connected Cartesian regions permit comprehensive
sampling, in principle.
However, since the sampling is discrete, and Jacobian can be illconditioned,
the issue of narrow gates at
unknown locations in
Cartesian regions needs to be dealt with.
Here we leverage the fact that Cayley space is convex.
The idea is to use previous Cayley step that stayed in feasible Cartesian region as a new step. 
We can guarantee that this will not reverse direction or repeat sample 
in Cartesian space.
In short, for every point close to the boundary in Cartesian,
we check if it is possible to walk on Cayley space. 




\begin{algorithm*}[h!tbp]  %
\SetKwInOut{Input}{input}\SetKwInOut{Output}{output}
{\bf jumpToDisconnectedRegion}\\
  \Input{$s_i$, $M$, cartpoint, config \tcp*[f]  {$M$:the matrix keeps
  track of a Cartesian point is visited or not, cartpoint: the
  vector for current Cartesian point in $M$} }
  \Output{new\_config } 
  \BlankLine
 	compute new\_config by walking from $config$ by step := $s_i$ \tcp*[f] this is a jump on Cartesian space\\
	\If{ new\_config stays in Cayley boundary} 
	{		
		compute new\_cartpoint of new\_config; \\
		\If { $M$( new\_cartpoint ) is not visited before }
		{
			\If { new\_cartpoint moves at least $1$ Cartesian step from cartpoint} 
			{
				return new\_config;
			}
		}		
	}
	return failed; \\
  \caption{jumpToDisconnectedRegion} 
\label{alg:jumpToDisconnectedRegion} 
\end{algorithm*}

\section{Results}
\label{experiments}

Recall that our goal is to combine the advantages of Cayley sampling with
that of uniform sampling in Cartesian. The former
permits topological roadmapping, as well as guaranteed isolation and 
coverage of effectively low dimensional, low potential energy
regions relatively much more efficiently and with much 
fewer samples compared to MonteCarlo or simply Cartesian 
grid sampling, with the additional efficiency of not leaving the feasible
regions, and not discarding samples.
\cite{Ozkan2011, Ozkan2014MainEasal}. 

\begin{figure}
\centering
 \begin{subfigure}{.2\textwidth}
\epsfig{file = 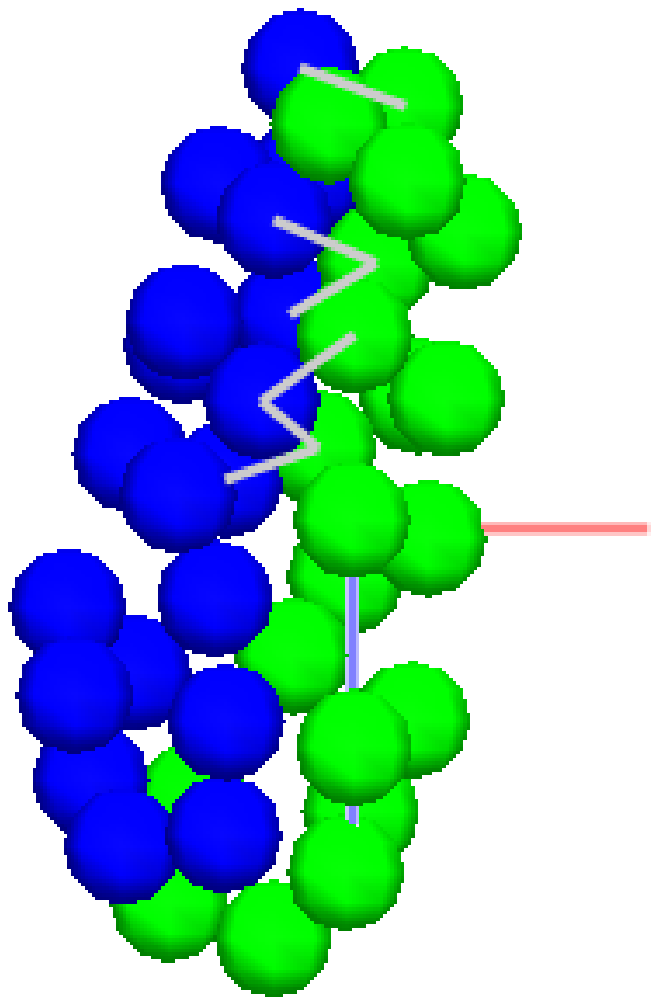, width = \linewidth }
\caption{HiaPP}
   \end{subfigure}
 \hskip0.01\linewidth
\begin{subfigure}{.18\textwidth}
\epsfig{file = 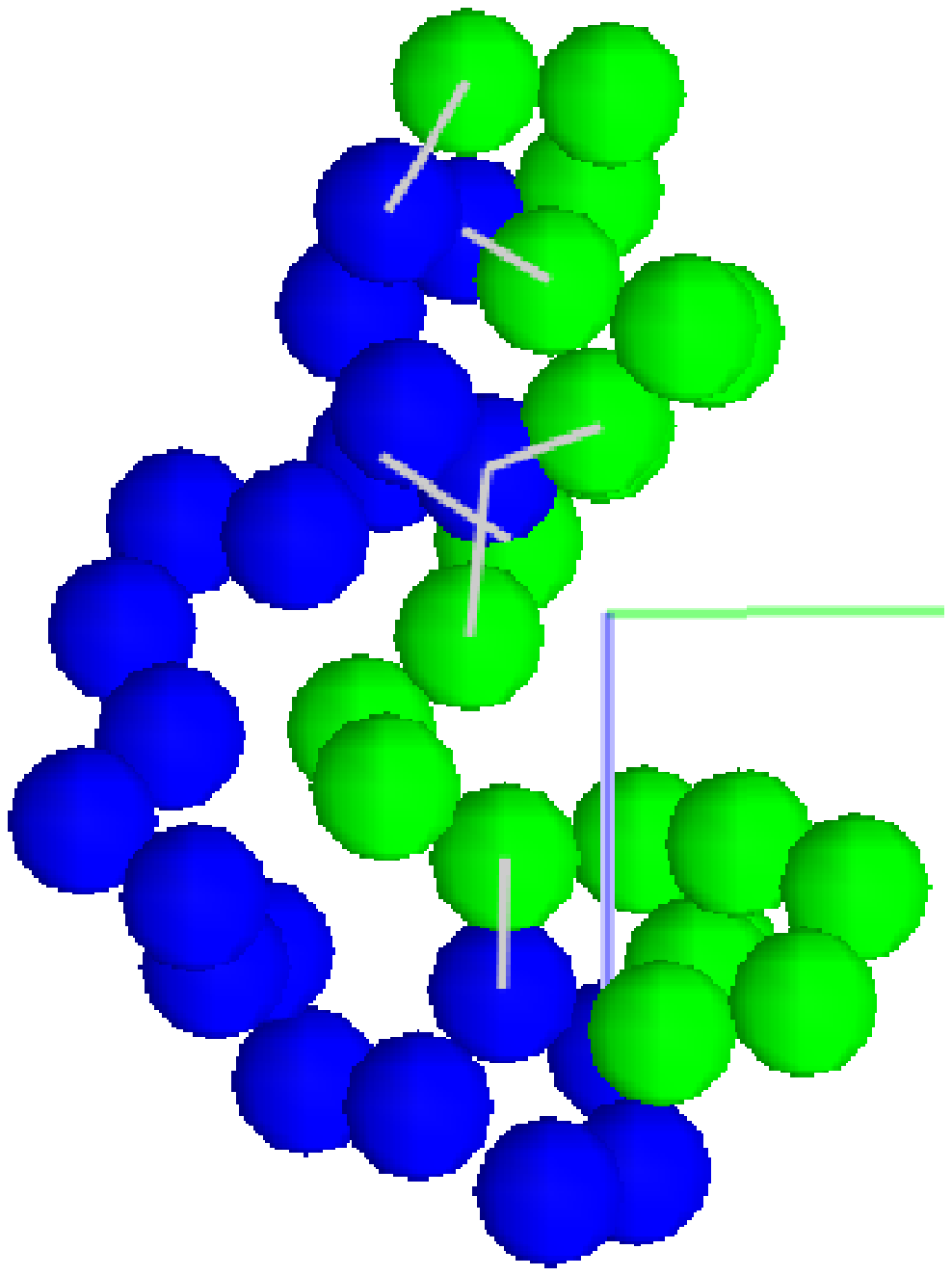, width = \linewidth }
\caption{RiaPP}
   \end{subfigure}
\caption{Easal screenshot: displays the molecule } 
\label{fig:pdb} 
\end{figure}

\emph{Since the methods of this paper have preserved the above advantages, 
the emphasis of our comparison here is only the
uniformity of sampling in Cartesian}. 
For this purpose only,  
we compare the original EASAL \cite{Ozkan2011, Ozkan2014MainEasal}, 
modified EASAL-jacobian
(this paper) and uniform Cartesian grid sampling
of assembly configuration spaces of $2$ rigid molecules with about $20$ atoms.
We used last 20 residues of HiaPP(human islet amyloid polypeptide
PDB-2KJ7) which contain the 6 residues where it differs from RiaPP (rat islet amyloid polypeptide PDB-2KB8). See fig.~\ref{fig:pdb}.
We created the 5D stratum (regions with a single active constraint) 
of both versions of EASAL atlas for $2$ assembling HiaPP molecules and
separately, for $2$ assembling RiaPP molecules.
For comparison purposes, in both cases, 
a reference Grid is generated, which is designed to
cover the part of the configurational space of interest, i.e., 
observed in nature.

\subsection{Multigrid}
Both versions of EASAL are designed to isolate and sample each active
constraint region.
In addition, EASAL-Jacobian samples each such region uniformly in Cartesian. 
Yet, when we combine all such regions, those regions where more pairs of 
atoms are in their Lennard-Jones wells (regions with more active constraints) 
will have denser sampling. i.e. 
EASAL tends to oversample the lower energy regions.
This is a positive feature of EASAL that we preserve in EASAL-Jacobian.
Since the 5D strata of the atlas generated by both versions of EASAL would 
sample a configuration that has
$l$ active constraints $l$ times (once for each of the 5D active constraint
regions in which the configuration lies), the meaningful comparison would
require similarly replicating such configurations in the grid, which we call
the \emph{multigrid}.

\subsection{Grid Generation}
\begin{itemize}
    \item
The Grid is uniform along the Cartesian configuration space. 
\item
The bounds of the Cartesian configuration space for both Grid and EASAL 
are:\\
$X, Y :$ -26 to 26 Angstroms\\
$Z :$ -7 to 7 Angstroms\\
\item
The angle parameters are described in Euler angles representation 
(Cardan angle ZXZ).\\
$\phi, \psi : -\pi$ to $\pi$
\item
Inter principal-axis angle $\theta < 30.0$ degrees where $\theta = a\cos(uv)$ where $u$ and $v$ are the principal axis of each rigid body. I.e. $u$ and $v$ are eigenvectors of the inertia matrix.
\item
Additionally, there is the pairwise distance lower bound criterion:\\
For all atom pairs $i,j$ belonging to different rigid molecular components, 
$d_{ij} > 0.8 ∗ (r_i + r_j )$ where $i$ and $j$ are residues, 
$d_{ij}$ is the distance for residues $i$ and $j$, 
$r_i$ and $r_j$ are the radius of residue atoms $i$ and $j$. \\
\item
147 Million grid configurations are generated in this manner.
\item
Over 93\% of them are discarded to ensure at least one pair $d_{ij} < r_1
+ r_2 + 0.9$, i.e, an active constraint and to eliminate collisions. About $9.6$ Million grid
configurations remain.
\end{itemize}

\subsection{Computational Time/Resources for EASAL}
The specification of the processor that EASAL executed is Intel Core 2 Quad CPU Q9450 @ 2.66GHz x 4 with Memory:3.9 GiB.\\
EASAL-Jacobian for input HIAPP took 2 days 9 hours 20 minutes(3440 minutes) and for input RIAPP took 3 days 14 hours 44 minutes(5204 minutes).\\
EASAL for input HIAPP took 5 hours 40 minutes(340 minutes) and for input RIAPP took 6 hours 52 minutes(412 minutes).\\

\subsection{Epsilon Coverage}
Ideally, we would expect each Grid point to be covered by at least one 
EASAL sample point that
is situated in an $\epsilon$-cube centered around a Grid point 
with a range of $2\epsilon$ in each of the 6 dimensions.

\begin{itemize}
    \item
The value of $\epsilon$ is computed as follows:
$\epsilon$ = ($\#$ of Grid points / \# of Easal points)$^ {1/6} / 2$
\item
We set  $\epsilon$  to be $\lceil \epsilon \rceil$  since grid points are by
definition a discrete number of steps from each other.
\item
In order to compute the coverage, we assign each EASAL sample to its closest
Grid point. Call those Grid points \emph{EASAL-mapped} Grid points. 
We say that a Grid point $p$ is \emph{covered} if there is at least one
EASAL-mapped Grid point within the $\epsilon$-cube centered around $p.$
\item
\underbar{$\epsilon$ for HiaPP:} 
The number of samples generated by Grid, EASAL and EASAL-jacobian were
9,619,435/194,595/2,861,926 respectively. The corresponding $\epsilon$ for
EASAL is $\lceil 49.4331^ {1/6} / 2 \rceil = \lceil 0.957869 \rceil $ and for
EASAL-Jacobian is $\lceil 3.36118^ {1/6} / 2 \rceil = \lceil 0.611954 \rceil $.
\item
\underbar{$\epsilon$ for RiaPP:} 
The number of samples generated by Grid, EASAL and EASAL-jacobian were 
13,267,314/319,016/4,744,878 respectively. The corresponding $\epsilon$ for
EASAL is  $\lceil 41.5882^ {1/6} / 2 \rceil = \lceil 0.930676 \rceil$ and for
EASAL-Jacobian is  $\lceil 2.79613^ {1/6} / 2 \rceil = \lceil 0.593467 \rceil$.
\end{itemize}

\subsection{Coverage Results}
The results show that \textbf{$96.21\%$} of Grid points are 
covered by EASAL-jacobian for HiaPP and  \textbf{$96.14\%$} of Grid points are 
covered for RiaPP. 

For basic EASAL, \textbf{$85.03\%$} of Grid points are 
covered for HiaPP and  \textbf{$85.46\%$} of Grid points are 
covered for RiaPP. 

Hence EASAL-jacobian is verified to have almost full coverage.

\subsection{Density Distribution}

\begin{figure*} [h!tbp]
\def\wid{.4\textwidth}
\centering
\begin{subfigure}[b]{\wid}
      \epsfig{file = 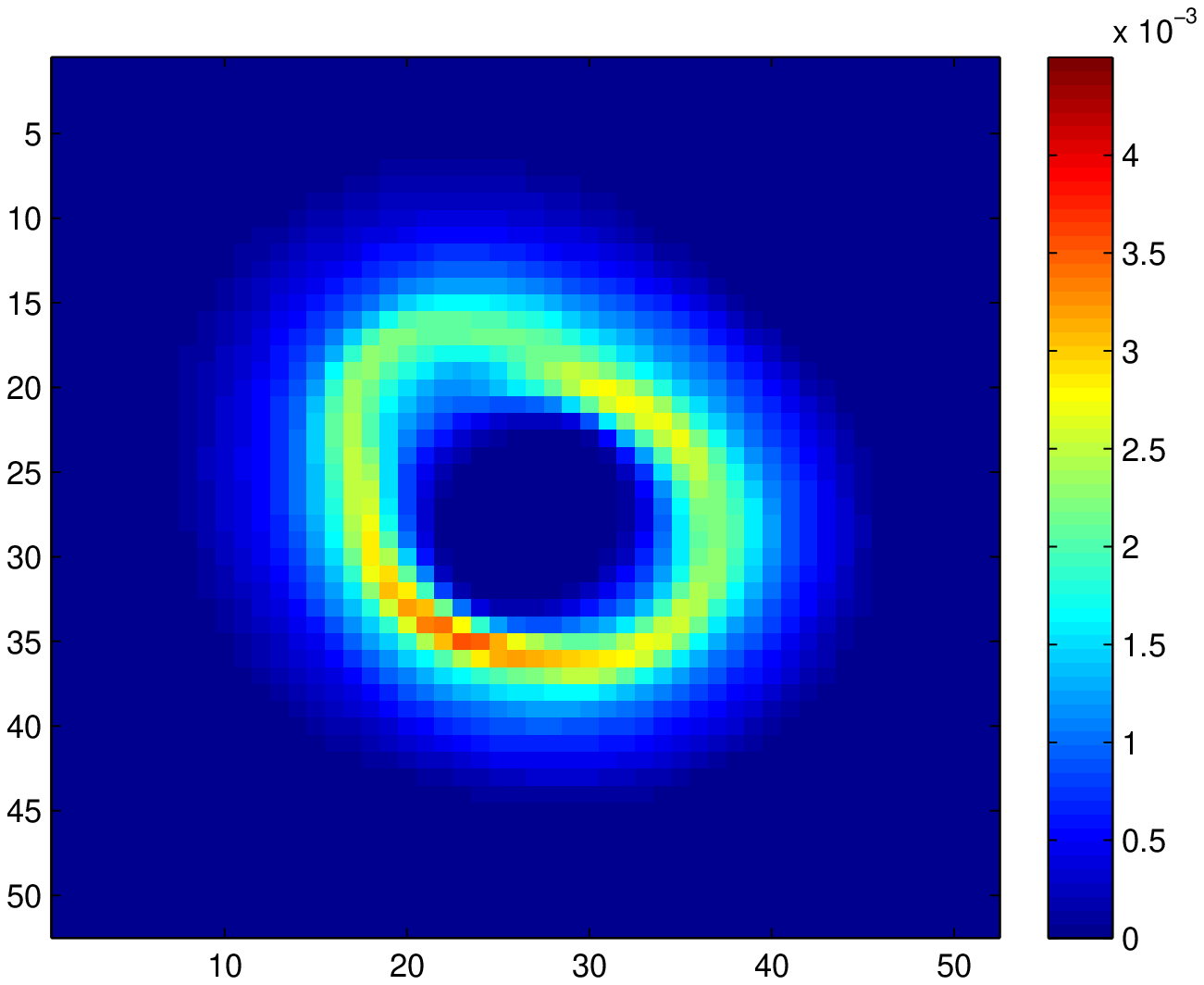, width=\linewidth}
      \caption{HiaPP: GRID}
   \end{subfigure}
\begin{subfigure}[b]{\wid}
      \epsfig{file = 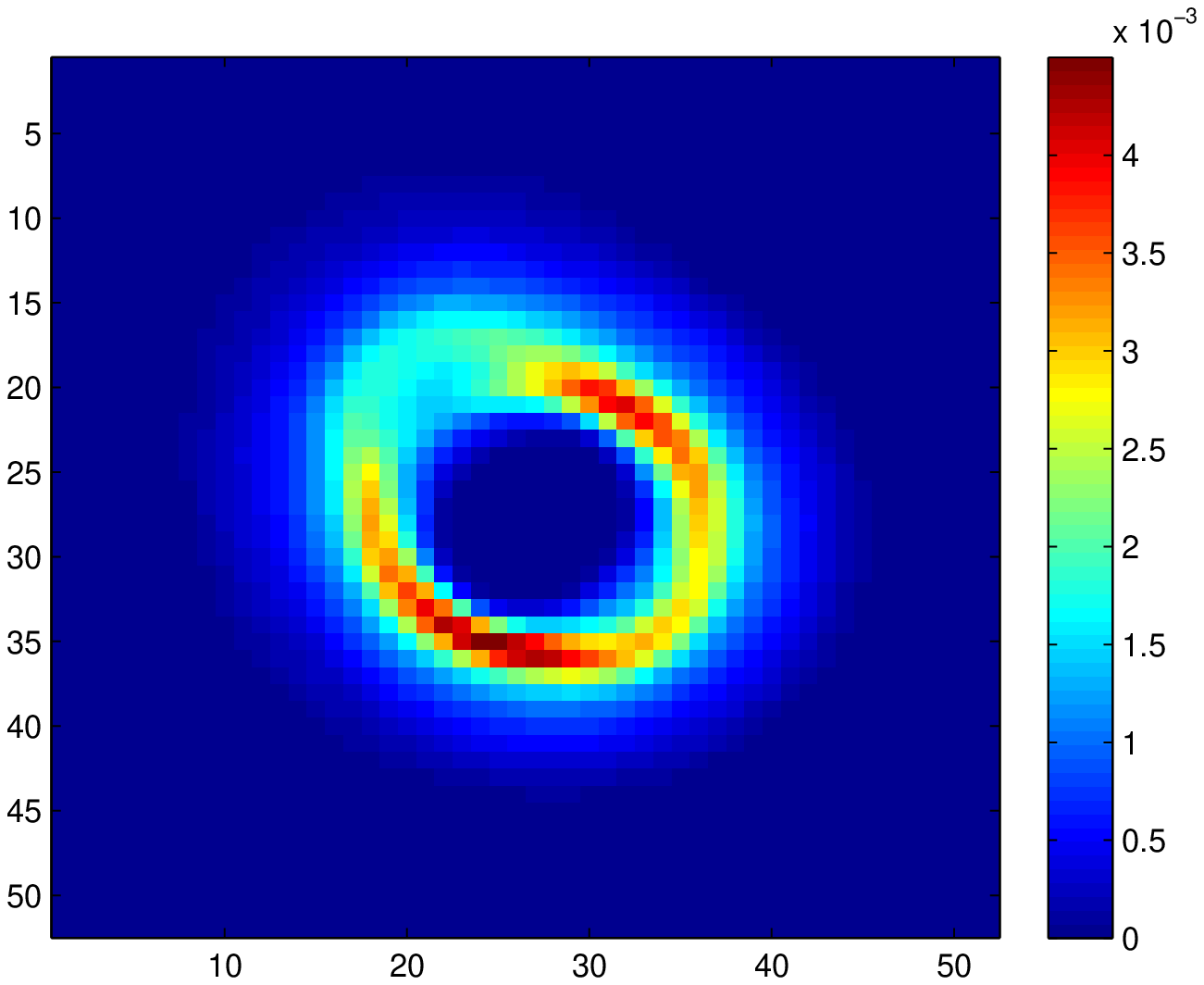, width=\linewidth}
      \caption{HiaPP: MULTIGRID}
   \end{subfigure}
 \begin{subfigure}[b]{\wid}
      \epsfig{file = 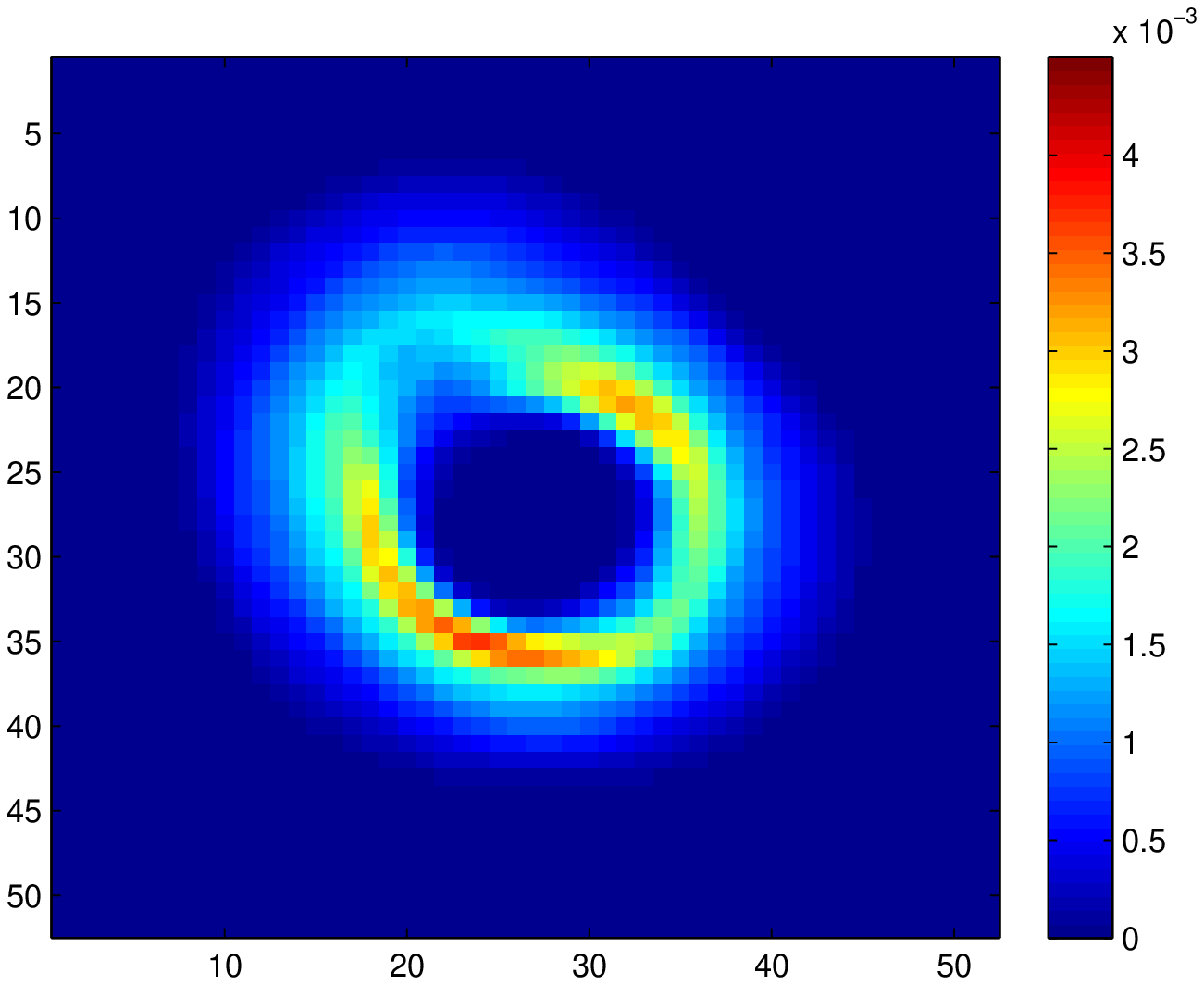, width=\linewidth}
      \caption{HiaPP: EASAL jacobian sampling}
   \end{subfigure}
\begin{subfigure}[b]{\wid}
      \epsfig{file = 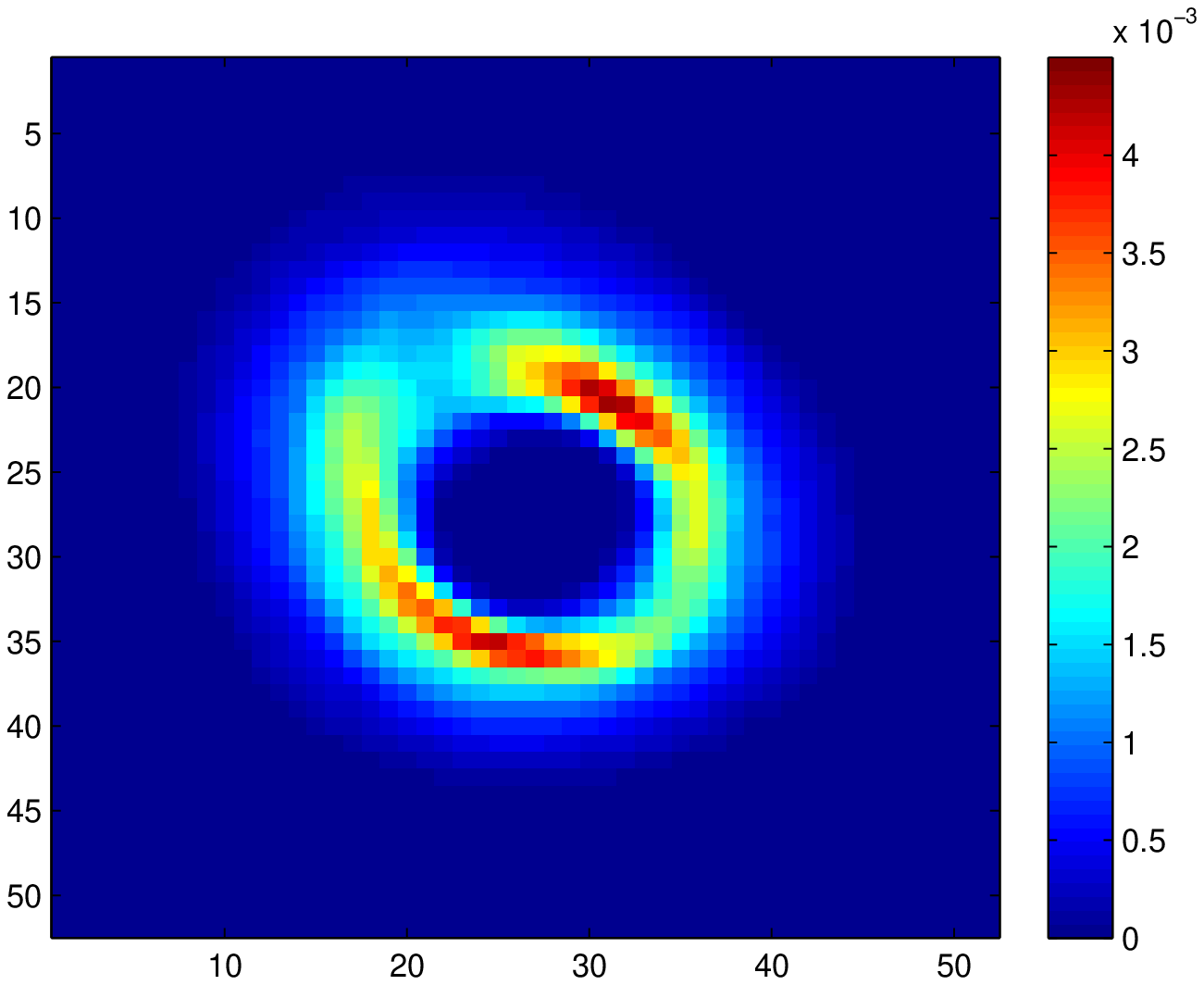, width=\linewidth}
      \caption{HiaPP: EASAL sampling}
   \end{subfigure}
 \begin{subfigure}[b]{\wid}
      \epsfig{file = 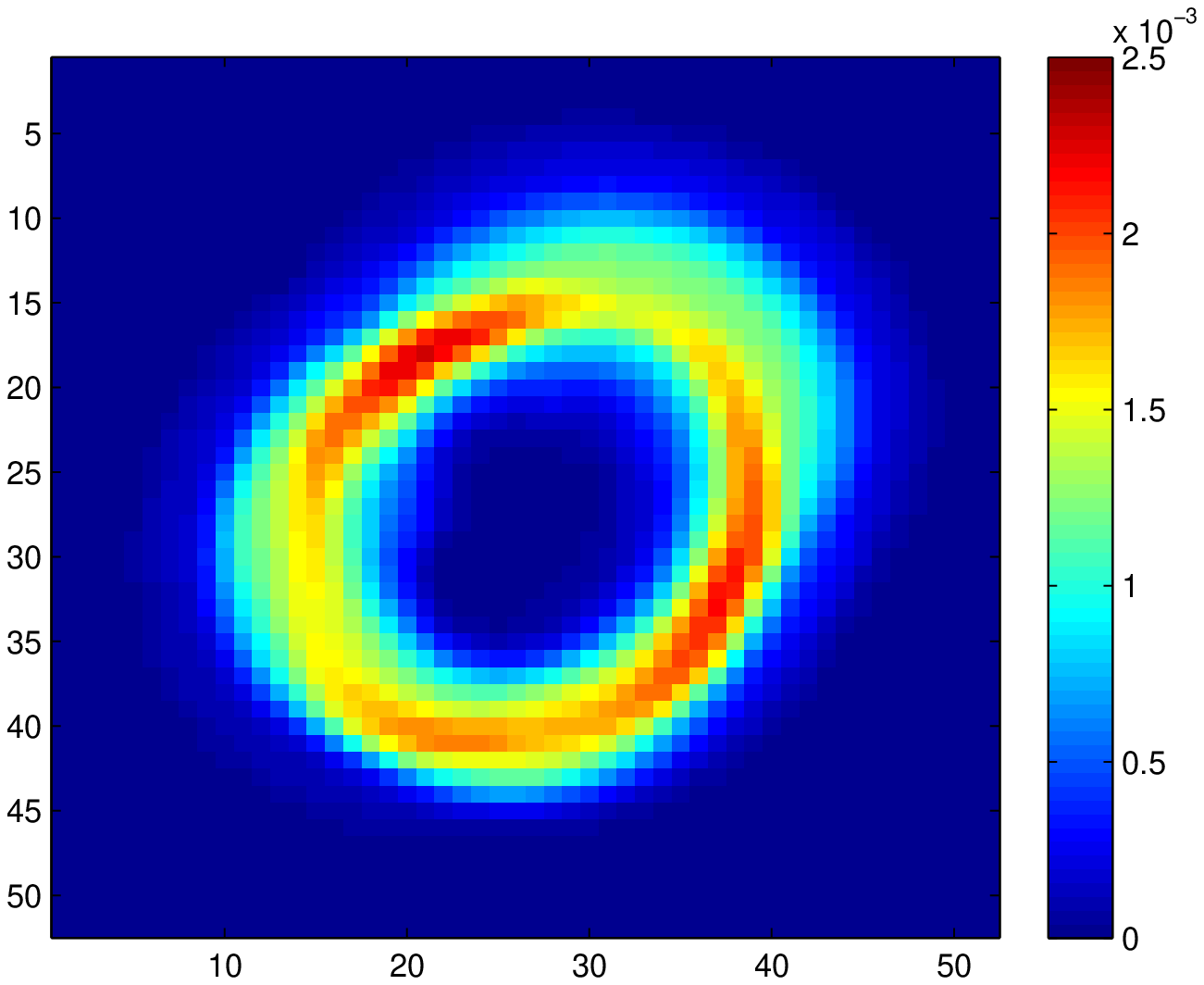, width=\linewidth}
      \caption{RiaPP: GRID}
   \end{subfigure}
 \begin{subfigure}[b]{\wid}
      \epsfig{file = 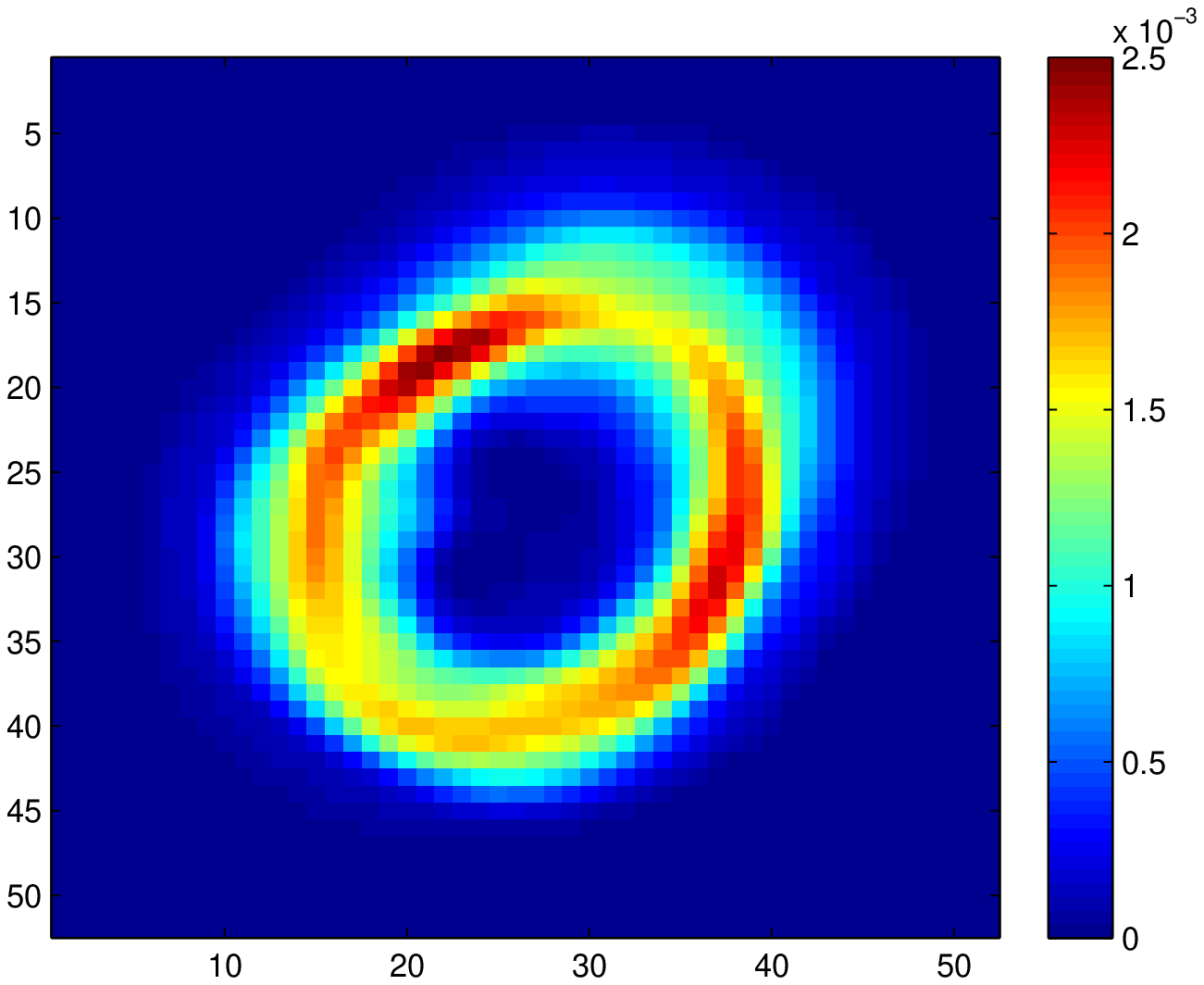, width=\linewidth}
      \caption{RiaPP: MULTIGRID}
   \end{subfigure}
\begin{subfigure}[b]{\wid}
      \epsfig{file = 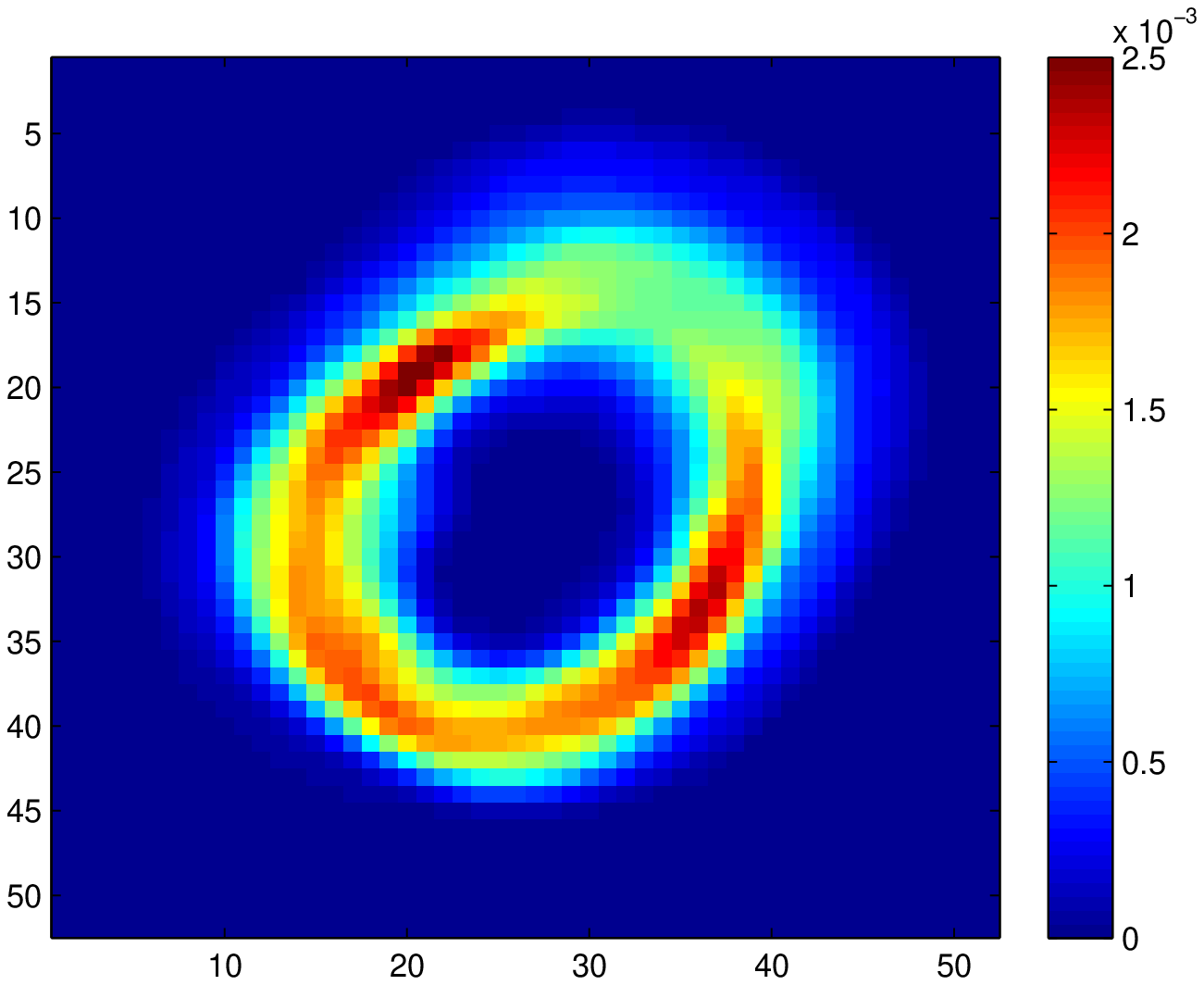, width=\linewidth}
      \caption{RiaPP: EASAL jacobian sampling}
   \end{subfigure}
\begin{subfigure}[b]{\wid}
      \epsfig{file = 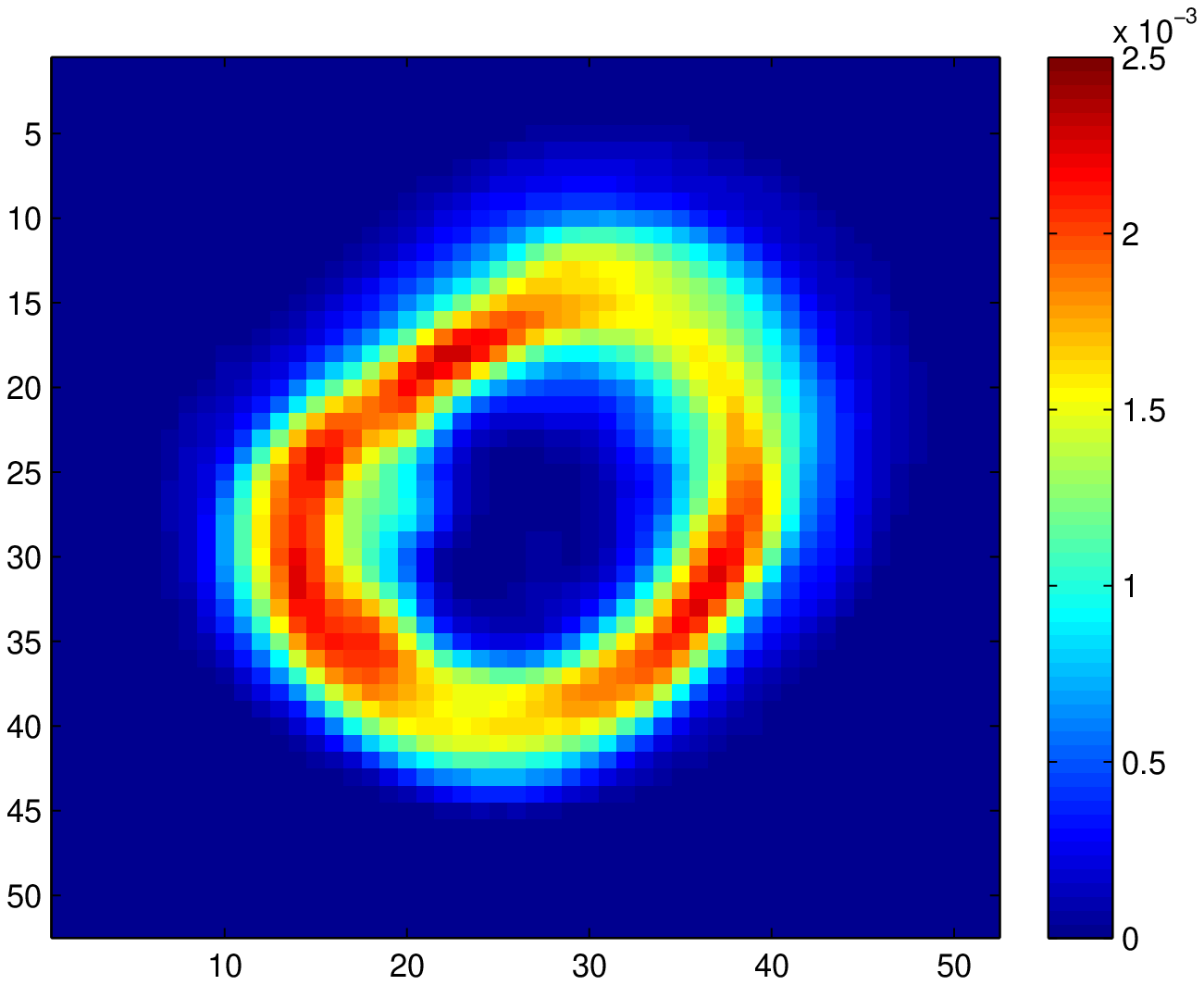, width=\linewidth}
      \caption{RiaPP: EASAL sampling}
   \end{subfigure}
\caption{ horizontal axis: Cartesian $x$ coordinate, vertical axis: Cartesian Y coordinate\\  
color code: the ratio of "the \# of points that lay in an $\epsilon$-cube
centered around Grid point $x,y$" over "total \# of points" \\
}
\label{fig:coverage_projection}
\end{figure*}

The fig.~\ref{fig:coverage_projection} shows the sampling distribution over
Cartesian $x,y$ space for Grid, MultiGrid, EASAL-jacobian and EASAL. The reddish regions are considered to be the lower energy regions.

EASAL and EASAL-jacobian is run for the majority of active constraint regions. i.e. it generated most of the 5D strata of the atlas.
Hence a configuration with $l$ active constraints is sampled close to $l$ times. Then we would expect density distribution for EASAL and EASAL-jacobian to lay in between Grid and MultiGrid. 

\section{Discussion}
A key goal is to find hybrid methods that combine the complementary
strengths of EASAL with prevailing methods.
A useful development would be a gradual tuning parameter, or 
flexible choice to allow a smooth transition from 
uniform sampling on Cartesian space to uniform sampling on Cayley space.
Such a tuning parameter would improve EASAL's flexibility  to go from 
the basic-EASAL to mimicking multigrid and MC, while still maintaining the 
advantages of EASAL. This would additionally make it easier to develop hybrids
between EASAL and prevailing methods leveraging the complementary advantages.
Extensive comparison of EASAL's and MC's performances have been reported in
\cite{Ozkan2014MC}.

Algorithm~\ref{alg:jumpToDisconnectedRegion} can be used with some
modifications as an independent component to improve ergodicity of 
regular MC sampling 
in  order to help jump to a region separated by a narrow
channel, or to pass a high energy barrier.  

Some aspects of the recursive and adaptive jacobian computation and 
sampling method presented here require a \emph{seed} matrix or direction
or value
starting from which they iterate. These include
Algorithms
\ref{alg:computeCayleyDirectionIteratively},
and \ref{alg:adaptiveMagnitudeAndDirection}.
In most cases, a good choice of seed is crucial for rapid convergence.

\section{Conclusion}
We have presented a modification to 
EASAL  that combines 
the advantages of Cayley sampling with
that of uniform sampling in Cartesian. The former
permits topological roadmapping, as well as guaranteed isolation and
coverage of effectively low dimensional, low potential energy
regions relatively much more efficiently and with much
fewer samples compared to MonteCarlo or simply Cartesian
grid sampling, with the additional efficiency of not leaving the feasible
regions, and not discarding samples.

The modification of EASAL presented here features careful and versatile 
use of the Jacobian of the maps between Cartesian and Cayley
to provide iterative, recursive and adaptive methods to achieve uniform
sampling in Cartesian while preserving the advantages of sampling 
in Cayley space.
While the results are encouraging when we compare the basic EASAL and the modified EASAL on
dimer assembly configuration spaces of HiaPP and RiaPP,
there is much room for  exploring and tapping the continuum of methods that 
traverse the distance between EASAL and 
other prevailing methods.


%
%

%



\bibliographystyle{plain}   
\bibliography{easal,pr,stickysphere,nigms,jorg,Dmay04}
\newpage

\end{document}